\documentclass[journal,twoside,web]{ieeecolor}
\usepackage{color,soul}
\usepackage{generic}
\usepackage{cite}
\usepackage{amsmath,amssymb,amsfonts}
\usepackage{algorithmic}
\usepackage{graphicx}
\usepackage{textcomp}
\usepackage{hyperref}
\usepackage{booktabs}
\usepackage{subcaption}
\usepackage{tabularx}
\usepackage{caption}
\usepackage{balance}
\usepackage{float}
\usepackage{mwe} 
\usepackage[counterclockwise, figuresright]{rotating}

\def\BibTeX{{\rm B\kern-.05em{\sc i\kern-.025em b}\kern-.08em
    T\kern-.1667em\lower.7ex\hbox{E}\kern-.125emX}}
\begin{document}
\title{Long-term Dependency for 3D Reconstruction of Freehand Ultrasound Without External Tracker}
\author{Qi Li, Ziyi Shen, Qian Li, Dean C. Barratt, Thomas Dowrick, Matthew J. Clarkson, Tom Vercauteren, and Yipeng Hu
\thanks{Qi Li, Ziyi Shen, Dean C. Barratt, Thomas Dowrick, Matthew J. Clarkson and Yipeng Hu
are with the UCL Centre for Medical Image Computing, and the Wellcome/EPSRC Centre for Interventional and Surgical Sciences, Department of Medical Physics and Biomedical Engineering, University College London, London WC1E 6BT, U.K. (e-mail: qi.li.21@ucl.ac.uk; ziyi-shen@ucl.ac.uk; d.barratt@ucl.ac.uk; t.dowrick@ucl.ac.uk; m.clarkson@ucl.ac.uk; yipeng.hu@ucl.ac.uk).}
\thanks{Qian Li is with State Key Laboratory of Robotics and System, Harbin Institute of Technology, Harbin, China, and also with the UCL Centre for Medical Image Computing, and the Wellcome/EPSRC Centre for Interventional and Surgical Sciences, Department of Medical Physics and Biomedical Engineering, University College
London, London WC1E 6BT, U.K. (e-mail: qian-li@ucl.ac.uk).}
\thanks{Tom Vercauteren is with School of Biomedical Engineering \& Imaging Sciences, King’s College London, London WC2R 2LS, U.K. (e-mail: tom.vercauteren@kcl.ac.uk).}}

\maketitle
\begin{abstract}
\emph{Objective:} Reconstructing freehand ultrasound in 3D without 
any external 
tracker has been a long-standing challenge in ultrasound-assisted procedures. 
We aim to define new ways of parameterising long-term dependencies, and evaluate the performance.
\emph{Methods:} First, long-term dependency is encoded by transformation positions within a frame sequence. This is achieved by combining a sequence model with a multi-transformation prediction. Second, two dependency factors are proposed, anatomical image content and scanning protocol, for contributing towards accurate reconstruction. Each factor is quantified experimentally by reducing respective training variances.
\emph{Results:} 1) The added long-term dependency up to 400 frames at 20 frames per second (fps) 
indeed improved reconstruction, with an up to 82.4\% lowered accumulated error, compared with the baseline performance. 
The improvement was found to be dependent on sequence length, transformation interval and scanning protocol and, unexpectedly, not on the use of recurrent networks with long-short term modules; 2) Decreasing either anatomical or protocol variance in training led to poorer reconstruction accuracy. Interestingly, greater performance was gained from representative protocol patterns, than from representative anatomical features.
\emph{Conclusion:} The proposed algorithm uses hyperparameter tuning to effectively utilise long-term dependency. The proposed dependency factors are of practical significance in collecting diverse training data, regulating scanning protocols and developing efficient networks.
\emph{Significance:} 
The proposed new methodology with publicly available volunteer data and code\protect\footnote{\url{https://github.com/ucl-candi/freehand}} for parametersing the long-term dependency, experimentally shown to be valid sources of performance improvement, which could potentially lead to better model development and practical optimisation of the reconstruction application. 

\end{abstract}

\section{Introduction}
\label{sec:introduction}
\IEEEPARstart{D}{etermining} the relative 3D spatial positions between ultrasound (US) images can recover 3D anatomy and pathology in these images. 
External spatial trackers such as those based on mechanical, optical and electromagnetic principles, enabled many clinical ultrasound applications. Removing the need for such external devices has attracted decades of research interest, in order to devise a more portable, accessible and cost-effective freehand ultrasound system, without being constrained by line-of-sight \cite{benjamin2020renal} or magnetic field interference \cite{chung2017freehand}, particularly preferable in surgical and interventional applications.

Speckle-induced correlation between near-overlapping images has been studied to align spatially close US frames \cite{chen1997determination}. Statistical image correlation was also investigated \cite{prager2003sensorless}, under the same assumption. 
The image slice separation is limited in this kind of approach, usually chosen to be 0.2 mm. These methods have focused on inherent correlation between images, independent of clinical applications with specific protocols and their intended anatomical content, for estimating spatial transformation from images. 

To improve upon these general approaches, recent data-driven deep learning-based methods are capable of learning ``global'' correlations for determining relative image locations. Indeed, recurrent neural networks (RNNs) \cite{miura2020localizing}, \cite{miura2021probe}, \cite{luo2021self}, and transformers \cite{ning2022spatial} have been proposed to model US frames as sequential data, with reported better spatial localisation. For example, Luo et al \cite{luo2021self} tested ConvLSTM on sequences with 90-120 frames and Miura et al \cite{miura2021pose} used ConvLSTM with sequences of 180 frames. This suggests that, even though these methods lack a physical basis, there is still an advantage to be gained from image frames that are spatially and temporally distant, i.e. further than a few neighbouring frames which may contain shared content or signals. This is referred to as the long-term dependency in this work\footnote{In other fields such as natural language processing, long-term dependency may refer to those with much longer distances, therefore is considered an application-dependent term.}. However, most of the aforementioned studies have incorporated other innovative contributions such as novel network training strategies \cite{ning2022spatial}, \cite{xie2021image}, and added prior knowledge \cite{luo2021self}. It is therefore unclear whether and how much the reported improvement originated from this long-term dependency. 

This work describes a new freehand US sequence encoding together with a multiple transformation prediction algorithm. The correlations within the input US sequence, those between a large number of output transformations, and the output dependency on the input sequence can be readily modelled as hyperparameters of RNNs or, perhaps more interestingly, feed-forward convolutional neural networks (CNNs). We show that a large margin of improvement reducing up to 74.5\% in final drift, due to including distant (up to 20s) past frames, was possible for specific applications.

To investigate factors that resulted in this improvement, two types of application-specific long-term dependencies are hypothesized, \textit{anatomical dependency} and \textit{protocol dependency}. That is, predicting spatial frame locations is considered dependent on \textit{a)} anatomical/pathological content in acquired images and \textit{b)} on pre-defined scanning paths, probe movements or orientation patterns during image acquisition by trained operators. This work utilises the proposed method to quantify the performance change due to altering these two factors independently.

We argue in this paper that a better understanding of the benefits of long-term dependency, by quantifying reconstruction accuracy as a function of the factorised dependency, is not only an interesting research topic but also practically important. Dependency (hyper-)parameters, defined in Section~\ref{sec:parametric-dependency}, are useful for choosing effective models, whilst identifying performance-gaining dependency factors may help generalise these gains, for example by optimising protocols, training cohort of data and the trade-off between computation cost and memory requirement from dedicated hardware. 

The sequence modelling method was summarised in a conference paper \cite{li2023trackerless}. This work has a different focus on further developing the methodology specifically for modelling and assessing long-term dependency, with different experiments using more than 5 times longer sequences. The added contributions include: 
1) A detailed description and motivation of the proposed input encoding and multiple transformation output method;
2) Presenting extensive experimental results for demonstrating the improvement from long-term dependency; 
3) Based on analysis of the long-term dependency, in terms of the defined hyperparameters and proposed dependency factors, a number of interesting conclusions are summarised, some of which are addressed with quantitative evidence for the first time; and
4) The code and volunteer data are made available for public access to ensure study reproducibility and further research. 

\section{Related work}
\label{sec:Related_work}

3D US reconstruction, a promising technique for ultrasound examination and ultrasound-guided intervention, has advantages over its 2D counterpart in many clinical scenarios, such as multi-modal registration\cite{lang2012multi}, musculoskeletal assessments\cite{huang2015correspondence,chen2015development}, volume visualisation and measurement\cite{guo2022ultrasound}. A large number of approaches has been proposed for 3D US reconstruction, which in this paper are studied using three categories: 1) scanning with 2D-array US probe, which can directly acquire 3D US volume\cite{light1998progress}; 2) mechanical scanning, which can efficiently reconstruct the 3D US volume by using motorized mechanical motor to move the US transducer along predefined trajectories\cite{mercier2005review}; 3) freehand 3D US scanning, which can reconstruct 3D US using spatial-temporal information of probe obtained by tracker or trackerless methods.

Despite the perceived flexibility and accessibility associated with freehand 3D US scanning, a spatial tracker (often external) is required which adds on cost and other logistic challenges, such as maintaining line-of-sight for optical trackers and avoiding interference for electromagnetic trackers. It has therefore been a strong research interest in developing trackerless freehand 3D US reconstruction, which may historically be further classified into non-deep-learning and deep-learning based methods.

Many popular non-deep-learning based freehand 3D US systems are based on utilising speckle patterns in US images\cite{chen1997determination}. Although some consider speckle impacts the quality of 3D US reconstruction with studies trying to suppress speckle to enhance tissue contrast\cite{huang2009speckle}, the correlation of speckle could be analysed, using statistical or machine learning approaches, to indicate the likelihood of relative positioning of nearby US frames, therefore to achieve tracking without external trackers. Gao et al\cite{gao2016wireless} proposed a wireless and sensorless 3D US imaging system that relied on adaptive speckle decorrelation curve to measure the motion of US probe along a single direction.  This study has demonstrated the feasibility of image based US probe tracking method on phantom and real-tissue data, although more work remains to be done for allowing much less unconstrained scanning protocol so they can be clinically useful. 

Deep-learning based approaches, featured with helpful representation ability, have been utilized for 3D US reconstruction\cite{prevost20183d}. For instance, Guo et al\cite{guo2020sensorless} proposed a deep contextual learning network (DCL-Net), a sequence modelling method with 3D convolutions, attention module, and a novel case-wise correlation loss, for 3D US reconstruction. Luo et al\cite{luo2022deep} exploited acceleration and orientation data measured by inertial measurement unit (IMU) to extract velocity information that could help estimate elevational displacements better. They also proposed an online self-supervised strategy for adaptive optimization of the model to reduce the drift. Based on this work, Luo et al afterwards proposed a multi-IMU-based network to reduce noise in IMU data\cite{luo2023multi}, in which a modal-level self-supervised strategy for IMU information fusion and a sequence-level self-consistency strategy for estimation stability enhancement were presented for performance improvement, demonstrated by extensive ablation experiments. A self-supervised learning and adversarial learning based online learning strategy was presented in\cite{luo2023recon}, along with a motion-weighted training strategy, for case-wise adaption to unseen dataset with diverse scanning velocities and poses. Instead of formulating a transformation into rotation and translation components separately, Hou et al\cite{hou2018computing} trained a pose estimation CNN on manifold $SE(3)$, with a left-invariant Riemannian metric. This proposed loss, computed on Riemannian geodesic space, could couple the translation and rotation components, taking into account the structure of Lie group $SE(3)$. Yeung et al proposed a pipeline for mapping 2D US images into 3D space with a pairwise comparison module and attention mechanism\cite{yeung2021learning}. Inspired by\cite{wang2021nerf}, Yeung et al parameterised the 3D reconstruction with implicit neural representation, jointly refining the initial pose estimation. A regression CNN was used in\cite{di2022deep} with the continuous rotation representation\cite{zhou2019continuity}, demonstrated on both phantom and real fetal data.
Wein et al\cite{wein2020three} proposed a pipeline for 3D thyroid assessment, consisting of tracking estimation, joint co-registration, and thyroid segmentation.

In summary, deep-learning-based trackerless freehand 3D US reconstruction seems a promising alternative to previous approaches without using deep neural networks. Existing methods estimated probe positions from US sequence that contained more than two frames, with the assumption that long-term dependency across the sequence can benefit estimation of current probe positions. However, to our best knowledge, no existing methods quantified such long-term dependency and analysed its contributing factors, the two aims of this study. The hypothesis of long-term dependency will be examined and defined in the following sections.

\section{Method}\label{Method}

Assume a sequence of 2D US frames $S=\{I_m\}, m=1,2,...,M$, with a sequence length $M$, and denote the spatial transformation between $i^{th}$ and $j^{th}$ frames as $T_{j\leftarrow i}, 1 \leq i<j \leq M$. In this work, $T_{j\leftarrow i}$ is represented by homogeneous matrices describing the relative translation and rotation, such that points $p^{(i)}$ in $i^{th}$ image coordinate system, in $[x,y,z,1]$ homogeneous coordinates, can be transformed to $j^{th}$ image coordinate system, $p^{(j)} = T_{j\leftarrow i} \cdot p^{(i)}$, thus describing the relative positions between the two frames.

\subsection{Input Ultrasound Sequence Modelling}\label{sec:sequence_modelling}

Recurrent models, such as RNNs with long short-term memory (LSTM) modules \cite{hochreiter1997long} and transformers \cite{vaswani2017attention} can be used to model the input sequential US frames. In this work, we assume a single intended output transformation\footnote{This assumption of single fixed output is made to enable the hyperparameters described in Section~\ref{sec:parametric-dependency} and for investigating image-to-transformation distance without predictions at multiple time points, further discussed in Section~\ref{sec:scan_reconstruction}.} $T_{j^*\leftarrow i^*}$ between two predefined frames $i^*$ and $j^*$, and a RNN model $f_{rnn}$, with network parameters $\theta$, is a function of individual frames $I_m$ at time step $m$ and the internal hidden state $h^{(m-1)}$ from the previous time step.
\begin{align} \label{eq:rnn}
    T_{j^*\leftarrow i^*} &=  f_{rnn}(I_m,h^{(m-1)};\theta), for~ m=M \nonumber \\
    h^{(m)} &= f_{rnn}(I_m,h^{(m-1)};\theta), \forall~m\leq{M-1}
\end{align}

This many-to-one mapping model enables the use of past frames $\{I_m\}_{m \in [1,i^*-1]}$ and future frames $\{I_m\}_{m \in [j^*+1,M]}$, the latter of which necessitates a temporal delay for a real-time system.

It is worth noting that the system or GPU memory required for training an unrolled $f_{rnn}$ is a function of the entire sequence, rather than individual frames, using back-propagation through time (BPTT) \cite{liao2018reviving}. Algorithms that are less dependent on sequence length, such as truncated BPTT or alternatives \cite{gruslys2016memory}, have seldom been seen in medical imaging applications, perhaps due to the potentially excessive computation for training high dimensional image input.

The recurrent models, with the single output at the end of a sequence, when unrolled, are conceptually equivalent to feed-forward models with the same output and the entire sequence as the input. This motivates us to test a CNN $f_{cnn}$ for this sequence modelling:
\begin{align} \label{eq:cnn}
T_{j^*\leftarrow i^*} &=  f_{cnn}(S;\theta)
\end{align}

\subsection{Output Multiple Transformation Prediction}\label{sec:MTL}

Although the sequence modelling described above only predicts a single transformation at the end of a sequence, supervision, i.e. ground-truth target transformations, at the previous time steps are available and were shown to accelerate training \cite{sangiorgio2020robustness}, also known as ``teacher forcing''. In this section, a multi-transformation prediction is proposed to use these additional data.

In addition to the intended $T_{j^*\leftarrow i^*}$, both the CNNs and RNNs can be adapted to output other $M(M-1)/2-1$\footnote{For an US sequence with length $M$, the possible number of output transformations can be $M$-combinations of a 2-set $C_M^2 = M(M-1)/2$.} transformations $\{T_{j\leftarrow i}\}, i\neq{i^*} or~ j\neq{j^*}$. Based on points $p^{(i)}_{n}$ sampled from $i^{th}$ frame in image coordinates, the proposed overall multi-transformation loss becomes:
\begin{equation} \label{eq:loss}
\mathcal{L}_{MTL} = \frac{1}{N \cdot M(M-1)/2} \sum^{M(M-1)/2}\sum_{n=1}^{N}\emph{D}_{mse}(p^{(j)}_{n},\hat{p}^{(j)}_{n})
\end{equation}
where $p^{(j)}_n$ and $\hat{p}^{(j)}_n$ are the same points transformed from the $i^{th}$ to $j^{th}$ image coordinate systems, $p^{(j)}_n = T^{(gt)}_{j\leftarrow i} \cdot p^{(i)}_{n}$ and $\hat{p}^{(j)}_n = \hat{T}_{j\leftarrow i} \cdot p^{(i)}_{n}$, using ground-truth $T^{(gt)}_{j\leftarrow i}$ and prediction $\hat{T}_{j\leftarrow i}$, respectively.
Four image corner points were used in this work, i.e. $N=4$.
Mean-square-error (MSE) was used as the distance function $\emph{D}_{mse}(\cdot)$, between $x$, $y$ and $z$ coordinates of the two points.

Optimising different transformations to the same image coordinate systems, as in Eq.~\ref{eq:loss}, encourages consistency and minimises accumulated error, as previously proposed \cite{li2023trackerless}. Using a third $k^{th}$ frame for example, when $\emph{D}_{mse}(T^{(gt)}_{j\leftarrow i} \cdot p^{(i)}_{n},\hat{T}_{j\leftarrow i} \cdot p^{(i)}_{n})$ is minimised simultaneously with $\emph{D}_{mse}(T^{(gt)}_{k\leftarrow i} \cdot p^{(i)}_{n},\hat{T}_{k\leftarrow i} \cdot p^{(i)}_{n})$ and $\emph{D}_{mse}(T^{(gt)}_{j\leftarrow k} \cdot p^{(k)}_{n},\hat{T}_{j\leftarrow k} \cdot p^{(k)}_{n})$, the difference between $\hat{T}_{j\leftarrow k} \cdot \hat{T}_{k\leftarrow i}$ and $T^{(gt)}_{j\leftarrow i}$ is thus implicitly minimised - a form of accumulated error, 
so is the difference between $\hat{T}_{j\leftarrow k} \cdot \hat{T}_{k\leftarrow i}$ and $\hat{T}_{j\leftarrow i}$, with equal ground-truth $T^{(gt)}_{j\leftarrow k} \cdot T^{(gt)}_{k\leftarrow i} = T^{(gt)}_{j\leftarrow i}$ - optimising a measure of consistency between sequential predictions.
As these multiple output transformations share information and impose regulating constraints on each other, each of them can be regarded as one task in a multi-task learning framework. The multi tasks consist of one main task $T_{j^*\leftarrow i^*}$ and other $M(M-1)/2-1$ auxiliary tasks. This multi-task learning framework takes advantage of the shared information among different tasks, which may result in improved performance over single-task learning framework. In addition, this multi-task learning framework can predict various transformations with various intervals, past and future frames using a common set of layers, making it possible to compare the performance of different hyperparameters in a single training run.
In practice, when $M(M-1)/2$ is large, $\tau+1$ transformation tasks, including various transformation intervals and number of past and future frames, are sampled due to memory limit, where $\tau \leq M(M-1)/2-1$.

\subsection{Parametric Dependency as Hyperparameters} \label{sec:parametric-dependency}
\begin{figure*}[h]%
\centering
\includegraphics[width=0.8\linewidth]{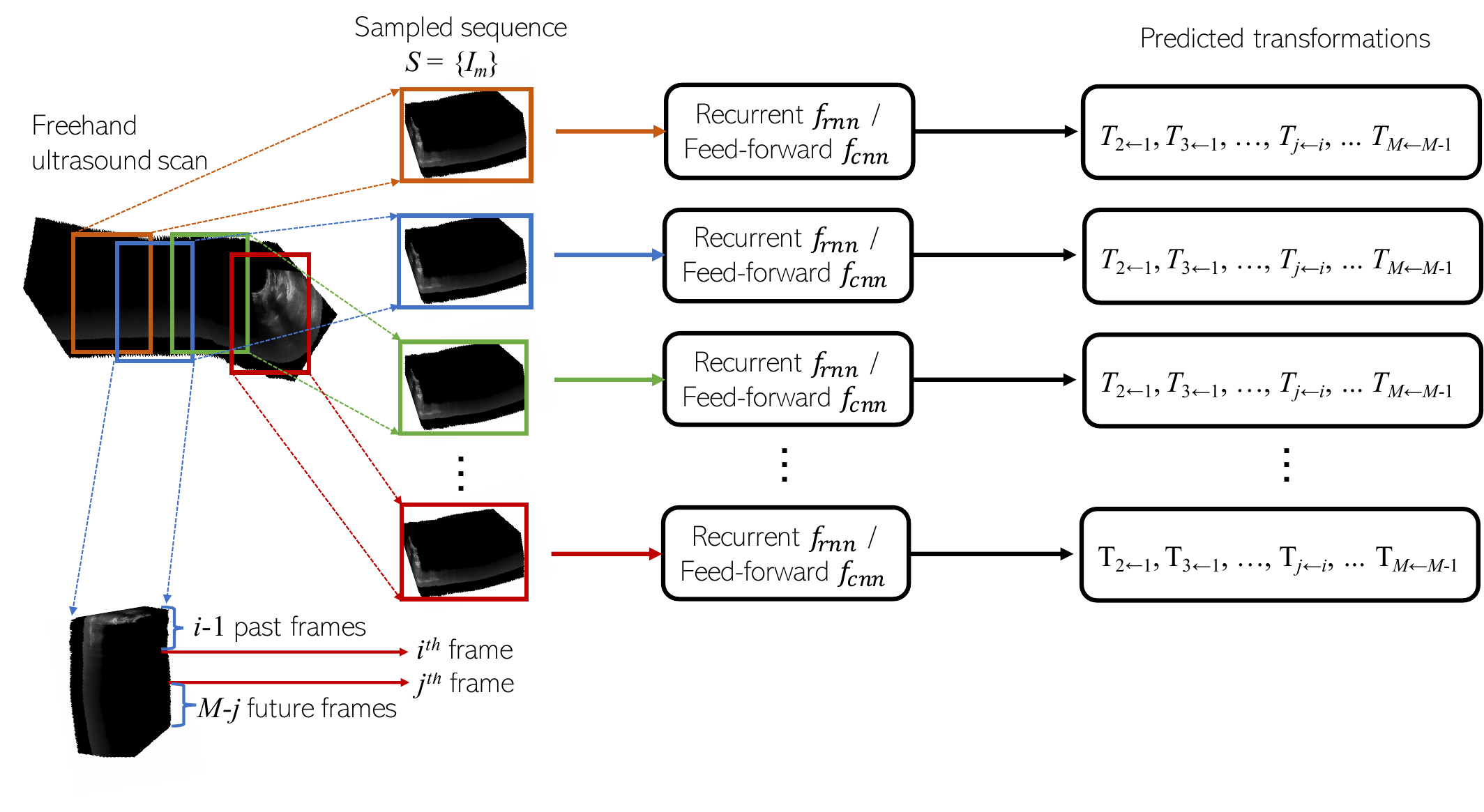}
\caption{Illustration of the proposed transformation-prediction algorithm.}\label{input_output}
\end{figure*}

As formulated in Sections~\ref{sec:sequence_modelling} and \ref{sec:MTL}, the dependency of transformation prediction can be quantified and illustrated in Fig.~\ref{input_output}. Past- and future- dependencies are represented by the number of the respective frames, $i^*-1$ and $M-j^*$. We propose to use $i^*$, $j^*$ and $M$ as hyperparameters of the models in Eqs.~\ref{eq:rnn} and \ref{eq:cnn}, which are general enough to represent many scenarios to test the dependency of the predicted transformation on frames outside of the transformation. For example, larger $i^*$ and $M-j^*$ increase the lengths of past  and future dependencies, respectively, whilst a high $M$ value can test both. It is noteworthy that tuning these hyperparameters may therefore aim for an optimum $T_{j^*\leftarrow i^*}$ on the validation set, rather than the overall loss in Eq.~\ref{eq:loss}. An extension to this work may investigate cases that predict a future or past transformation using acquired frames before or after the input sequence.

Sequence length $M$ determines the largest possible number of past and future frames that can be used for predicting transformations. For example, more past and future frames can be used with a larger $M$ and a smaller transformation interval. $M$ can be selected based on specific applications. For this study, the relationship between sampled and tested sequence lengths and reconstruction performance is reported in Section~\ref{sec:results_dependency}.

\subsection{Sequence Sampling and 3D Scan Reconstruction}\label{sec:scan_reconstruction}
From available US scans with variable lengths, sequences $S=\{I_m\}$ with the predefined $M$ are randomly sampled, for training models in Eqs.~\ref{eq:rnn} and \ref{eq:cnn}. The ground-truth is used to transform points in the $i^{th}$ image coordinates to the $j^{th}$ image coordinates, 
$T^{(gt)}_{j\leftarrow i} = T_{(calib)}^{-1} \cdot (T^{(gt)}_{world\leftarrow j})^{-1} \cdot T^{(gt)}_{world\leftarrow i} \cdot T_{(calib)}$,
where $T^{(gt)}_{world\leftarrow j}$ and $T^{(gt)}_{world\leftarrow i}$ are $j^{th}$-tool-to-world and $i^{th}$-tool-to-world transformations, at the time steps $j$ and $i$, obtained from the optical tracker. Thus, the transformation is independent of the world coordinate system. $T_{(calib)}$ is a fixed transformation from image to tool coordinate systems, obtained through spatial calibration. In practice, the left-multiplying inverse calibration matrix is not used, to which the loss is invariant to, due to the distance preservation property of orthogonal matrix. Thus, the loss is computed in the $j^{th}$ tracking tool coordinate system with a unit of millimeter (mm): $T^{(gt)}_{j\leftarrow i} = (T^{(gt)}_{world\leftarrow j})^{-1} \cdot T^{(gt)}_{world\leftarrow i} \cdot T_{(calib)}$. 

During the test, a scan can be reconstructed by predicting the optimum  $T_{j^*\leftarrow i^*}$ from consecutive sequences, such that the $(j^*)^{th}$ frame from the previous sequence is the $(i^*)^{th}$ frame in the subsequent sequence. Depending on the application where localising every possible adjacent frames is required, varying starting reference frames can be used and the relative locations between them may be determined by the auxiliary tasks, an independent initialisation method or potentially fixed with a predefined protocol. Furthermore, advanced hyperparameter selection methods, adaptive at different time points, and model ensembles to combine different predictions (therefore multiple main tasks) at the same time point, may further optimise the 3D reconstruction. These remain research interests for future study and are not considered in this work.

\subsection{Evaluation Metrics}\label{sec:Evaluation_metrics}

As the training loss is computed by using prediction error on each frame, one direct model generalisation metric \textit{frame prediction accuracy} ($\epsilon_{frame}$) is computed, as the difference between prediction $\hat{p}^{(j)}_{n}$ and ground-truth points $p^{(j)}_{n}$ on the $j^{th}$ frame, both transformed from the $i^{th}$ frame, $\epsilon_{frame} = \frac{1}{N}\sum_{n=1}^{N}\emph{D}_{dist}(p^{(j)}_{n},\hat{p}^{(j)}_{n})$,
where $\emph{D}_{dist}(\cdot)$ denotes the Euclidean distance between two points and $N=4$ on four corner points. This metric is useful for monitoring training and model development, but may not be indicative of the performance in predicting $T_{j^*\leftarrow i^*}$ or scan reconstruction.

For each test scan\footnote{This study was performed in accordance with the ethical standards in the 1964 Declaration of Helsinki and its later amendments or comparable ethical standards. Approval was granted by the Ethics Committee of local institution (UCL Department of Medical Physics and Biomedical Engineering) on $20^{th}$ Jan. 2023 [24055/001].}, three metrics are reported to assess the reconstructed frames: 1) \textit{Accumulated tracking error} ($\epsilon_{acc.}$) is the average Euclidean distance of all reconstructed image pixels between prediction and ground-truth,
$\epsilon_{acc.} = \frac{1}{\mathcal{J}\cdot N}\sum_{j=1}^{\mathcal{J}}\sum_{n=1}^{N}\emph{D}_{dist}(p^{(j)}_{n},\hat{p}^{(j)}_{n})$,
where $N$ is the number of pixels in an image and $\mathcal{J}$ is the number of reconstructed frames using $\hat{T}_{j\leftarrow i}$; 2) \textit{Volume reconstruction overlap} ($\epsilon_{dice}$) is a Dice score, computed as the overlap between the ground-truth- and prediction- reconstructed scan volumes; and 3) \textit{Final drift} ($\epsilon_{drift}$) measures as the average Euclidean distance, over the four corner points on the last frame of the scan, between ground-truth and prediction. 

\subsection{Factorised Dependency and Reduced Variance Analysis} \label{sec:va_reduced}

Anatomy dependency refers to the long-term dependency contributed by anatomical self-correlation with respect to anatomical variance. The common spatial movement pattern inherent in scanning protocols can also contribute to the long-term dependency, defined as protocol dependency. The abundance of anatomy and scanning protocol within the training dataset determines how much long-term dependency exists and can be learned. As US scan acquired from the same subject have the same anatomical content, anatomy dependency can be investigated by altering the included number of subjects. In the volunteer study, the dataset mentioned in Section \ref{sec:experiments} are acquired by three types of scanning protocols - straight line shape, ``C'' shape, and ``S'' shape, as variance of scanning protocol is determined by number of scanning protocols involved.
For quantifying anatomical dependency, the original training set can be re-sampled, at the subject level, such that a percentage of $v_{a}=25\%,50\%,75\%$ of the subjects are randomly removed, on which the same networks are trained and subsequently tested on the same test data. The difference in performance is quantified with respect to the reduced variance. For investigating the performance changes due to reduced protocol dependency, two sets of additional models are trained, `straight' only and `c-shape and s-shape', using one and two from the three different types of scans. Together with the models trained on all three types of scans, they represent three levels of protocol variance, $v_{p}=1,2,3$. Anatomical and protocol variances may both relate to various percentage of frames in a scan, denoted as $v_{l}$, $v_{l}=50\%,75\%$, and can be tested using training scans that are cropped to 50\% and 75\% of their original lengths.  

\section{Experiments} \label{sec:experiments}
\subsection{Data Acquisition}\label{Data_acquisition}

US data were acquired on an Ultrasonix machine (BK, Europe) with a curvilinear probe (4DC7-3/40), from 19 volunteers on both their left and right forearms. Three trajectories, straight, c-shape and s-shape, in a distal-to-proximal direction, were acquired for each forearm. For each trajectory, two scans were obtained by keeping the US probe approximately perpendicular of and parallel to the scanning direction, as illustrated in Fig. {\ref{scan_tra}.
Thus, six US scans were acquired on each forearm, each scan containing $36-430$ frames ($100-200$ mm). 
The dataset contains 228 scans in total, with statistics summarised in Fig. \ref{statistics_dataset}, and was split into train, validation and test sets by a ratio of 3:1:1 on the scan level.
Images with a size of 480×640 were recorded at 20 fps. 
The frequency was fixed at 6 MHz with a depth of 9 cm, a dynamic range of 83 dB, an overall gain of 48\%, and the speckle reduction was set at median level and the persistence at 3. Spatial calibration from image to tool coordinates was based on a pinhead-based method \cite{hu2017freehand}, and the temporal difference between the optical tracker (NDI Polaris Vicra, Northern Digital Inc., Canada) and imaging was calibrated using the Plus Toolkit~\cite{lasso2014plus}. 

\begin{figure}[h]%
\centering
\includegraphics[width=\linewidth]{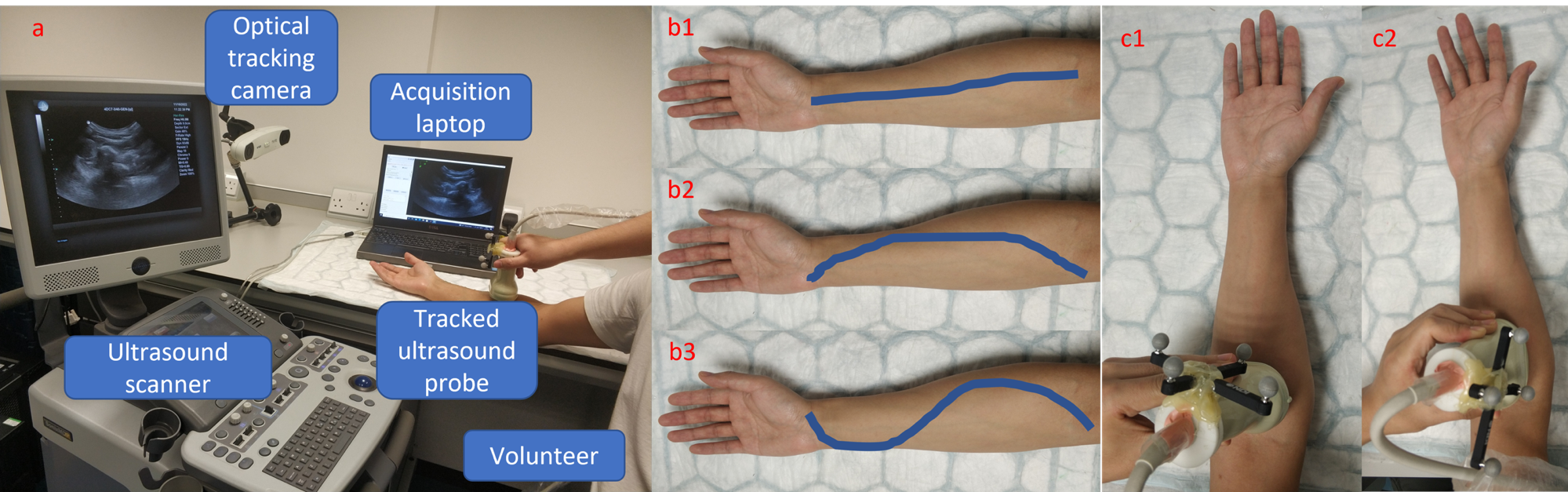}
\caption{Photographs of a) the US data acquisition system, b) various US probe trajectories, and c) various US plane orientations.}\label{scan_tra}
\end{figure}

\begin{figure}[h!]
        \centering

        \begin{subfigure}[t]{0.46\linewidth}
            \includegraphics[width=\linewidth]{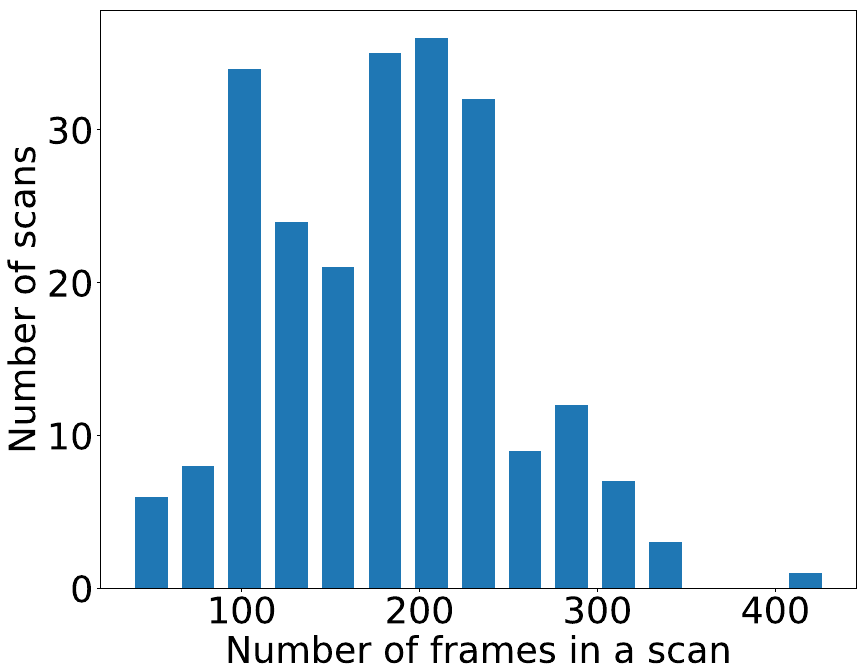}
            \caption[]%
            {{\small }}    
            \label{Histogram}
        \end{subfigure}
        \begin{subfigure}[t]{0.52\linewidth}
            \includegraphics[width=\linewidth]{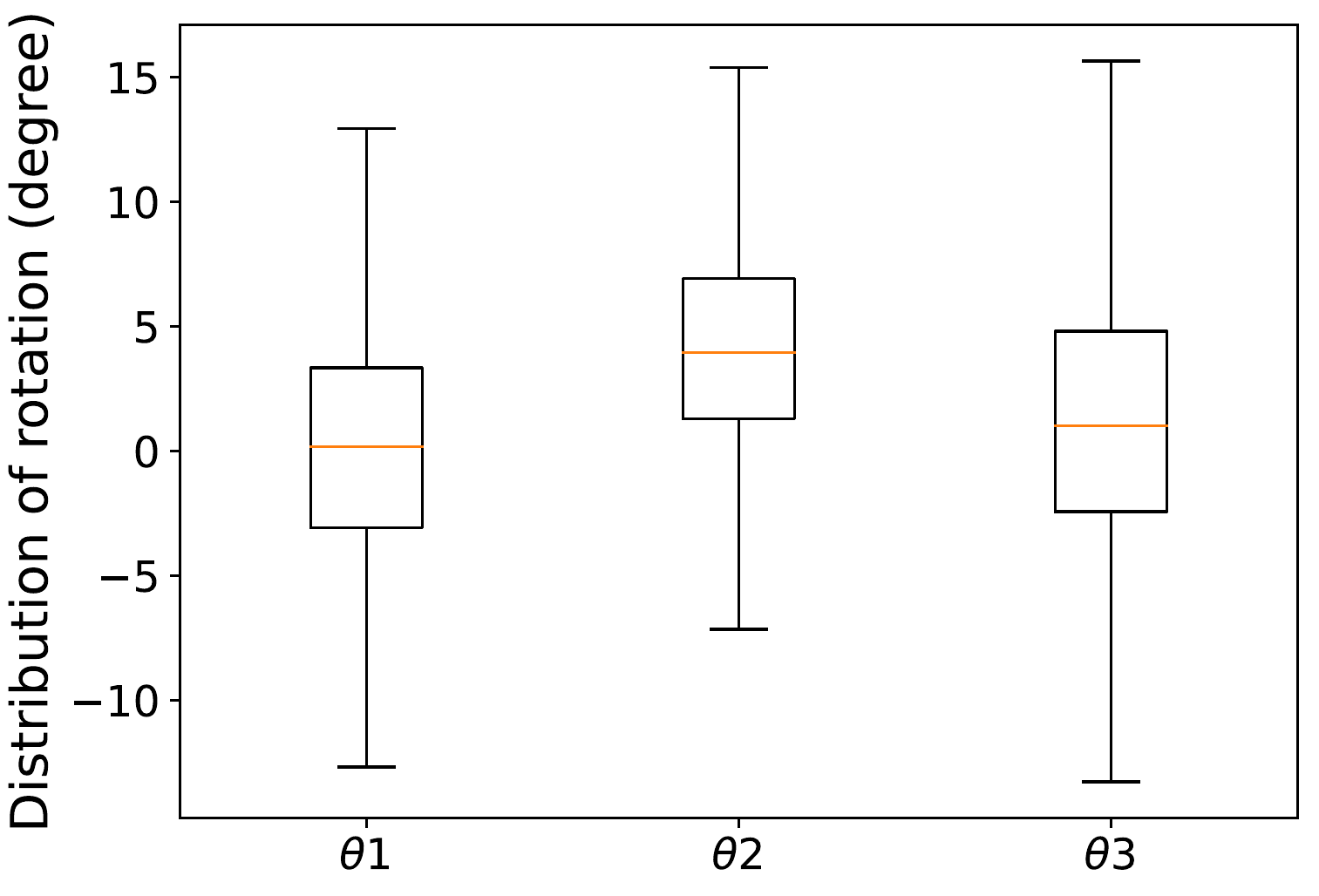}
            \caption[]%
            {{\small }}    
            \label{rotation_range}
        \end{subfigure}
        
        \hfill
        
        \begin{subfigure}[t]{0.49\linewidth}
            \centering
            \includegraphics[width=\linewidth]{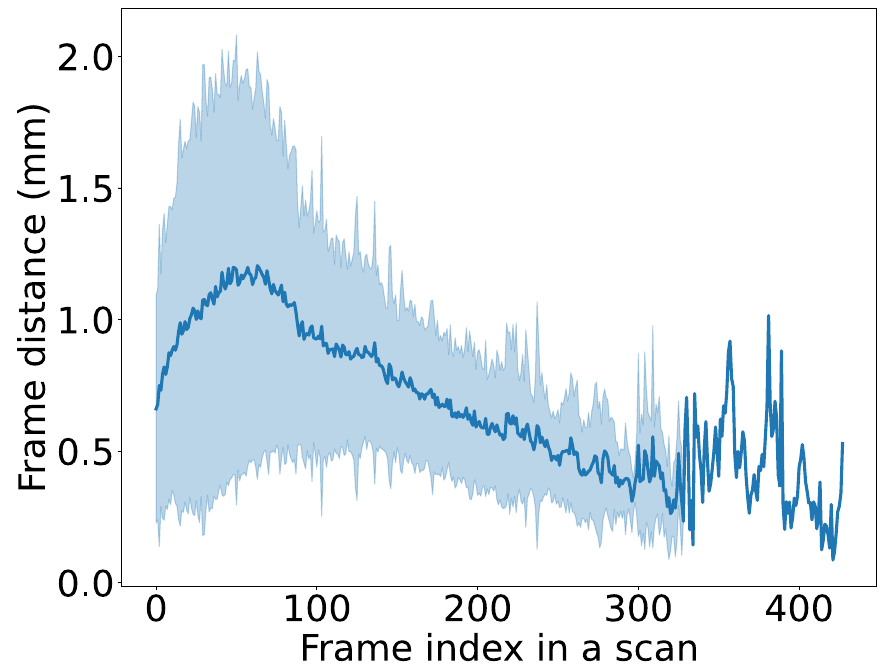}
            \caption[]%
            {{\small }}    
            \label{non_accumulated_dist}
        \end{subfigure}
        \hfill
        \begin{subfigure}[t]{0.49\linewidth}  
            \centering 
            \includegraphics[width=\linewidth]{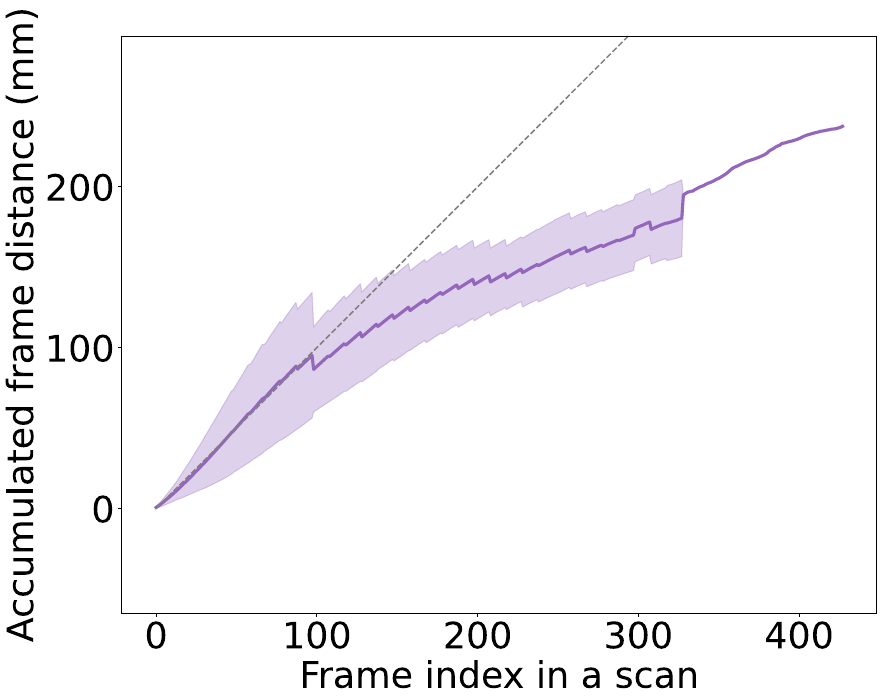}
            \caption[]%
            {{\small }}    
            \label{accumulated_dist}
        \end{subfigure}
        
        \caption[The histogram of scan lengths, mean and standard deviation of frame distance and accumulated frame distance of the dataset.]
        {Statistics of the dataset: (a) The histogram of scan length. (b) The distribution of three rotation parameters, between two adjacent frames. (c) Mean and standard deviation of distances between consecutive frames. 
        (d) Mean and standard deviation of accumulated consecutive frame distances.
        } 
        \label{statistics_dataset}
\end{figure}

\subsection{Network Development and Implementation} \label{model_development}

Both CNN- and RNN- based networks were trained using randomly-selected $(\tau+1)$ tasks with variable $M$ sequence length, in order to quantify the impact on performance due to varying long-term dependency, as discussed in Section~\ref{sec:parametric-dependency}. 
The commonly used and well-established CNN- and RNN- based networks were adapted in this paper, without excessive fine-tuning, to benchmark the results. The EfficientNet (b1)\cite{tan2019efficientnet} was adapted as a CNN, outputting $(\tau+1) \times 6$ transformation parameters using fully connected layer, where $\tau$ denotes the number of auxiliary tasks. EfficientNet has the advantage of being smaller and more efficient than other CNN networks while still preserving state-of-the-art performance. The RNN architecture we used is LSTM with 1024-dimensional hidden states, for its capacity to capture long-term dependency, utilizing the same EfficientNet (b1) network as feature encoder (1000-dimensional feature vectors).

In this work, models were trained with $M=10, 20, 30, 40, 49, 60, 75, 100$ separately, with $\tau+1=45, 80, 124, 157, 165, 177, 197, 218$ sampled tasks\footnote{As the number of tasks is one of the major sources of computational complexity and memory consumption, we have selected and tested a subset of all possible transformation predictions, with various transformation intervals, and past and future frame numbers.}. This results in 16 RNN/CNN networks.  
To study the effect of long-term dependency on reconstruction accuracy, the same CNN and RNN models were adapted with input sequence length $M=2$ (i.e. only using two adjacent US frames as input and outputting the transformation between them), used as the baseline. The motivation is to investigate the effect of long-term dependency, with and without further input frames. A recent method for freehand 3D US reconstruction, DCL-Net\cite{guo2020sensorless} has also been implemented for comparison.
In addition, 4 models with re-sampled $M=20, 49, 75, 100$ were trained using the same strategy to quantify the performance change due to altered dependency, i.e. reduced anatomical and protocol variance, detailed in Section~\ref{sec:va_reduced}.

For all networks, a minibatch size of 32, Adam optimizer, and learning rate $10^{-4}$ were used. The minibatch size and optimizer were empirically selected based on validation set performance, and the learning rate $10^{-4}$ were tested among $\{10^{-3}, 10^{-4}, 10^{-5}\}$. To test the dependency hyperparameters, results from varying $M$, $i^*$ and $j^*$ were computed. Each network was trained for at least 20,000 epochs until convergence, for up to 9 and 4 days on Ubuntu 18.04.6 LTS with a single NVIDIA Quadro P5000 GPU card, for RNNs and CNNs, respectively. All the results are reported on the test set unless otherwise specified.

The optimum predicted transformations, evaluated on the validation set, will be regarded as the main task. However, which one is optimum is unknown before training the model. Therefore, equal weights are given to different tasks to ensure a fair opportunity for each potential main task. In addition, tuning the explicit weighting between tasks may be partially redundant given the large number of auxiliary tasks and different configurations when varying the three hyperparameters, $M$, $i^*$ and $j^*$, some of which are equivalent to weighting the tasks differently. For example, less auxiliary tasks correspond to a bigger weight on the main task, and vice versa.

\section{Results} \label{sec:results}

\subsection{Results with Varying Dependency Hyperparameters\protect\footnote{Other reconstruction metrics yielded the same conclusions as summarised above, and these detailed results are provided in Supplementary Material for brevity.}} \label{sec:results_dependency}

Fig.~\ref{cnn_lstm_acc_err_pf_ff} summarises the performance of $\epsilon_{acc.}$ with respect to past- and future- frames, from all the models, with all available training data, described in Section~\ref{sec:experiments}. The reconstruction error decreases when more past frames are used. For example, using 74 past frames, the CNN and RNN achieved $\epsilon_{acc.}$ of $9.44\pm0.50$ mm and $10.04\pm0.56$ mm, respectively. Both represent statistically significant improvement ($p$-value $<0.001$), based on unpaired t-test at a significance level at $\alpha=0.05$, compared with that from the baseline (i.e., $M=2$, $\epsilon_{acc.}=22.75$ mm) and DCL-Net ($\epsilon_{acc.}=22.15$ mm). The influence of past frames is further illustrated in Figs.~\ref{acc_err_m20-s} to~\ref{acc_err_m100-s} in Supplementary Material, showing that $\epsilon_{acc.}$ decreased with more past frames, when the number of future frames is fixed. Fig.~\ref{reconstruction_results2} also illustrates better reconstruction from longer sequence length.

Other interesting observations include: 1) When using fewer than 25 added past frames, the improvement is not obvious. For example, with 20 past frames, no statistically significant difference was found between the baseline and either CNN ($p$-value=0.762) or RNN ($p$-value=0.815); 2) There was no statistically significant difference found between CNN and RNN, which may suggest that feed-forward models are equally competent in modelling US sequences with limited length, compared to the more ``specialised'' RNNs; 3) The same trend was not found when future frames increased. This was first suspected to be caused by non-constant scanning speed, as illustrated in Fig. \ref{statistics_dataset} (d). Additional experiments are illustrated in Fig. \ref{acc_err_sample_speed_reverse_CNN} to investigate the relationship between the reconstruction performance and scanning speed. The two models are trained using re-sampled train data with relatively constant speed and a reversed frame order. No evidence shows the correlation between $\epsilon_{acc.}$ and the scanning speed.

Table~\ref{cnn_all_train_best} shows the effect of sequence length on reconstruction performance, using the best CNN models among sampled tasks, evaluated by four evaluation metrics. It can be concluded that larger $M$ generally results in a smaller reconstruction error, due to the utilization of a relatively larger long-term dependency, i.e. more past and future frames. On the other hand, larger $M$ corresponds to higher computational complexity, reflected in speed of forward/backward process and model convergence.

Fig. \ref{validation_loss} shows the train and validation losses, trained with the varying train sets. A relatively larger number (20,000) was empirically selected as the maximum training epoch to train the model until convergence. Although validation loss begins to increase after a relatively rapid decrease, the best model used during inference stage are selected based on the performance on the validation set.
Although the aim of this work is primarily to analyse the hyperparameters, consistent results on the validation set were also obtained, shown as in Fig. \ref{validation_loss}. This suggested the feasibility to tune these hyperparameters for optimum reconstruction for specific applications, on available validation sets.
\begin{figure}[h!]
        \centering
        \begin{subfigure}[b]{0.49\linewidth}
            \centering
            \includegraphics[width=\linewidth]{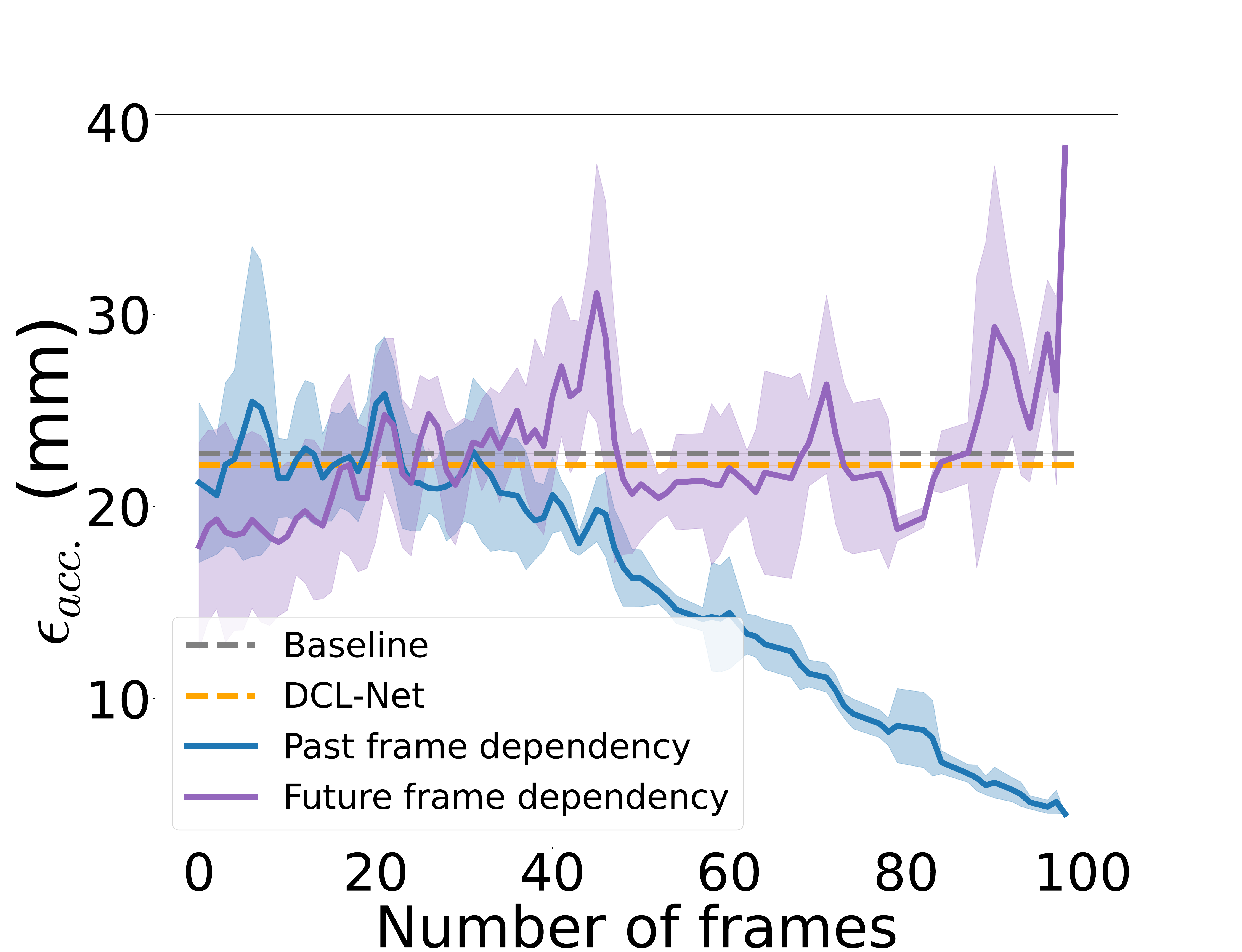}
            \caption[]%
            {{\small }}    
            \label{cnn_acc_err_pf_ff}
        \end{subfigure}
        \begin{subfigure}[b]{0.49\linewidth}  
            \centering 
            \includegraphics[width=\linewidth]{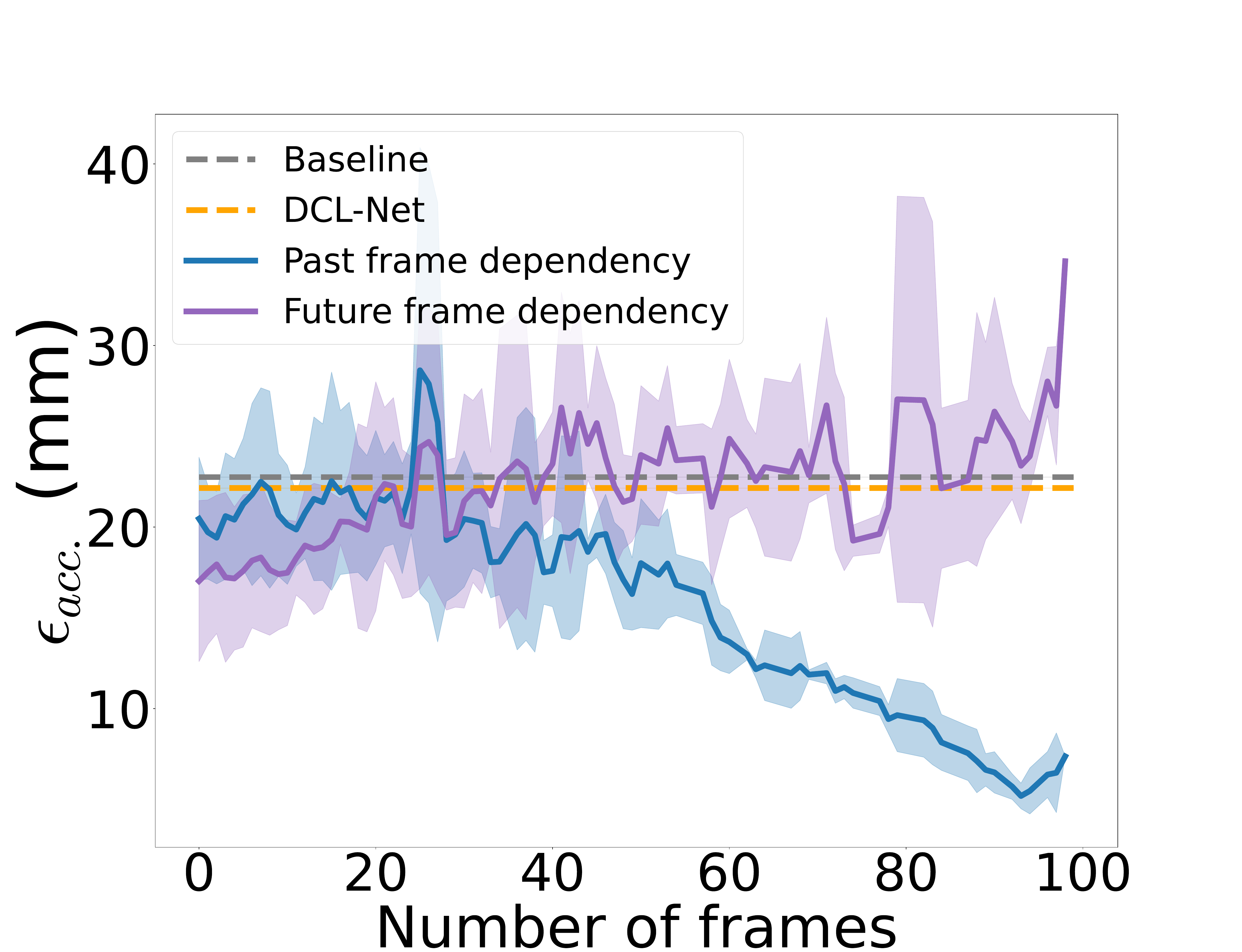}
            \caption[]%
            {{\small }}    
            \label{lstm_acc_err_pf_ff}
        \end{subfigure}
        
        \caption[ $\epsilon_{acc.}$ with respect to past- and future-dependency.]
        {$\epsilon_{acc.}$ with respect to dependency, from CNN (a) and RNN (b). The means and standard deviations of $\epsilon_{acc.}$ are plotted over all test scans, from models with $M=20, 49, 75, 100$.} 
        \label{cnn_lstm_acc_err_pf_ff}
\end{figure}

\begin{figure*}[h!]%
\centering
\includegraphics[width=0.9\linewidth]{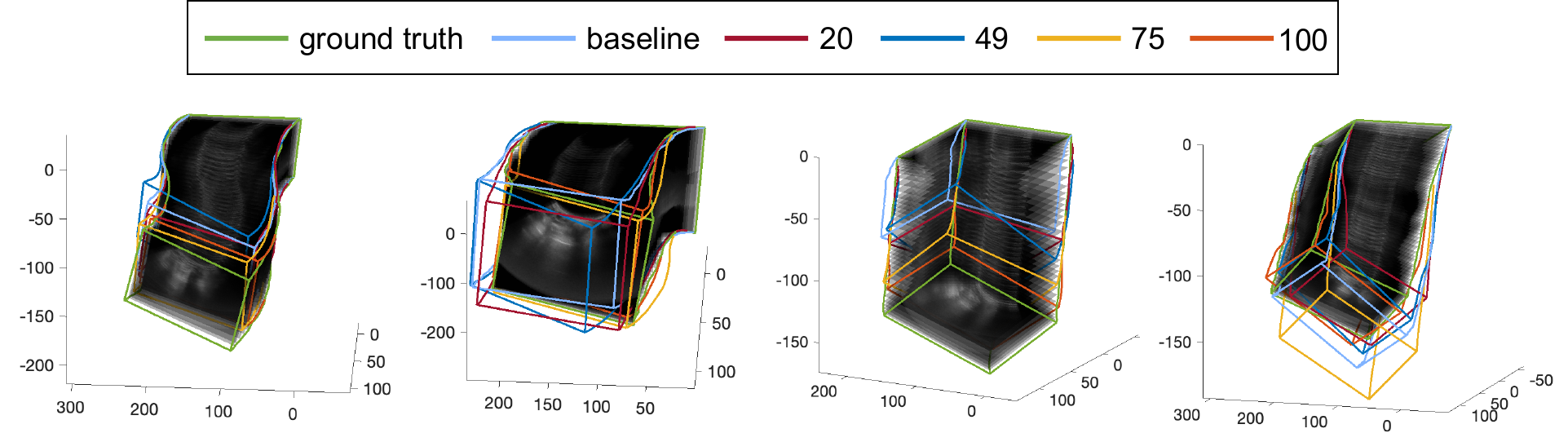}
\caption{Reconstruction from baseline and various $M=20,49,75,100$, with a selected transformation of $T_{18\leftarrow 8}$, $T_{32\leftarrow 30}$, $T_{69\leftarrow 64}$, and $T_{94\leftarrow 92}$, respectively. The trajectories are `S' shape, `C' shape, straight line shape, and straight line shape, from left to right, respectively.}\label{reconstruction_results2}
\end{figure*}

\begin{figure}[h!]
        \centering
        \begin{subfigure}[b]{0.49\linewidth}
            \centering
            \includegraphics[width=\linewidth]{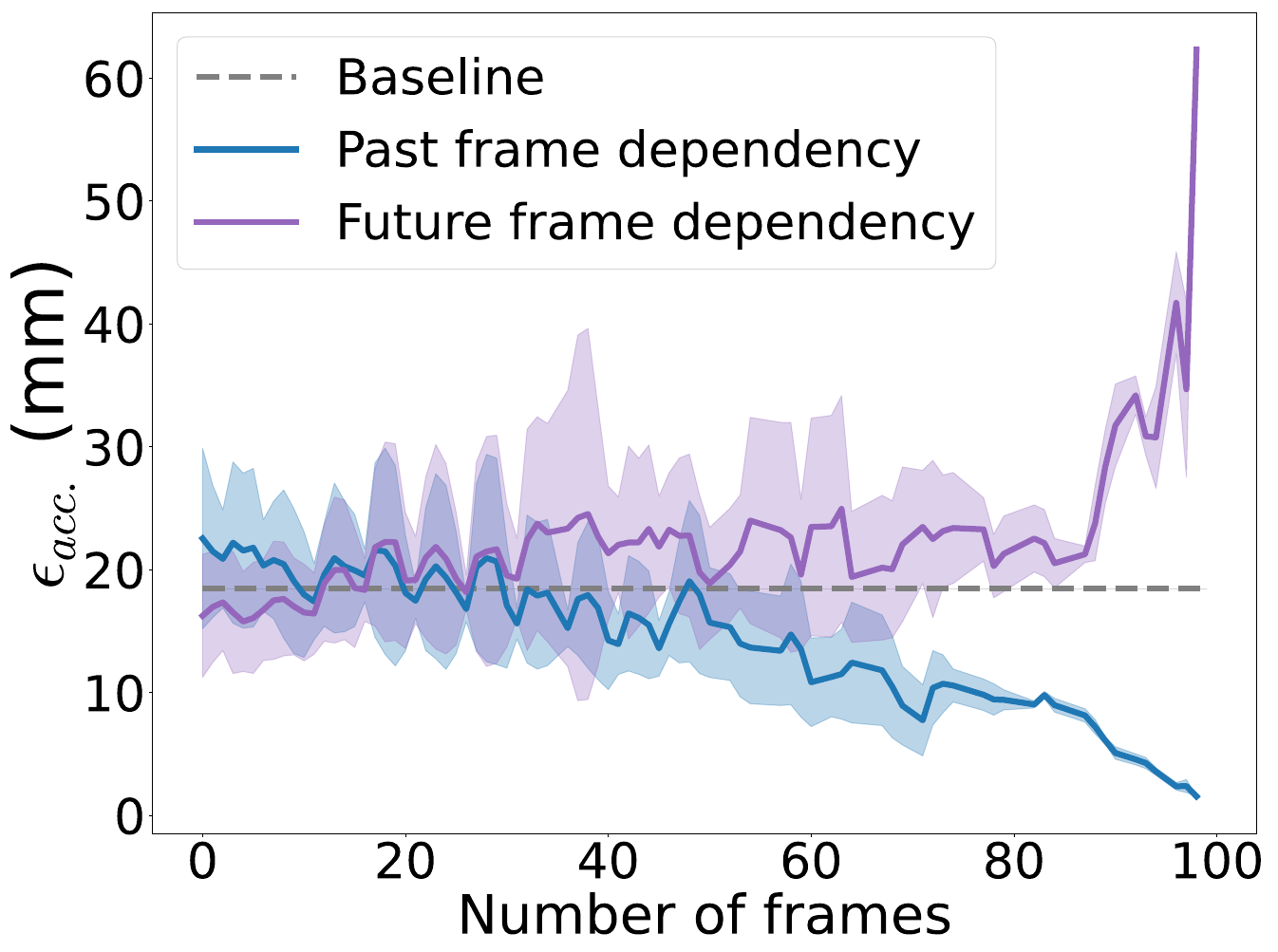}
            \caption[Network2]%
            {{\small}}    
            \label{acc_err_sample_speed_CNN}
        \end{subfigure}
        \hfill
        \begin{subfigure}[b]{0.49\linewidth}  
            \centering 
            \includegraphics[width=\linewidth]{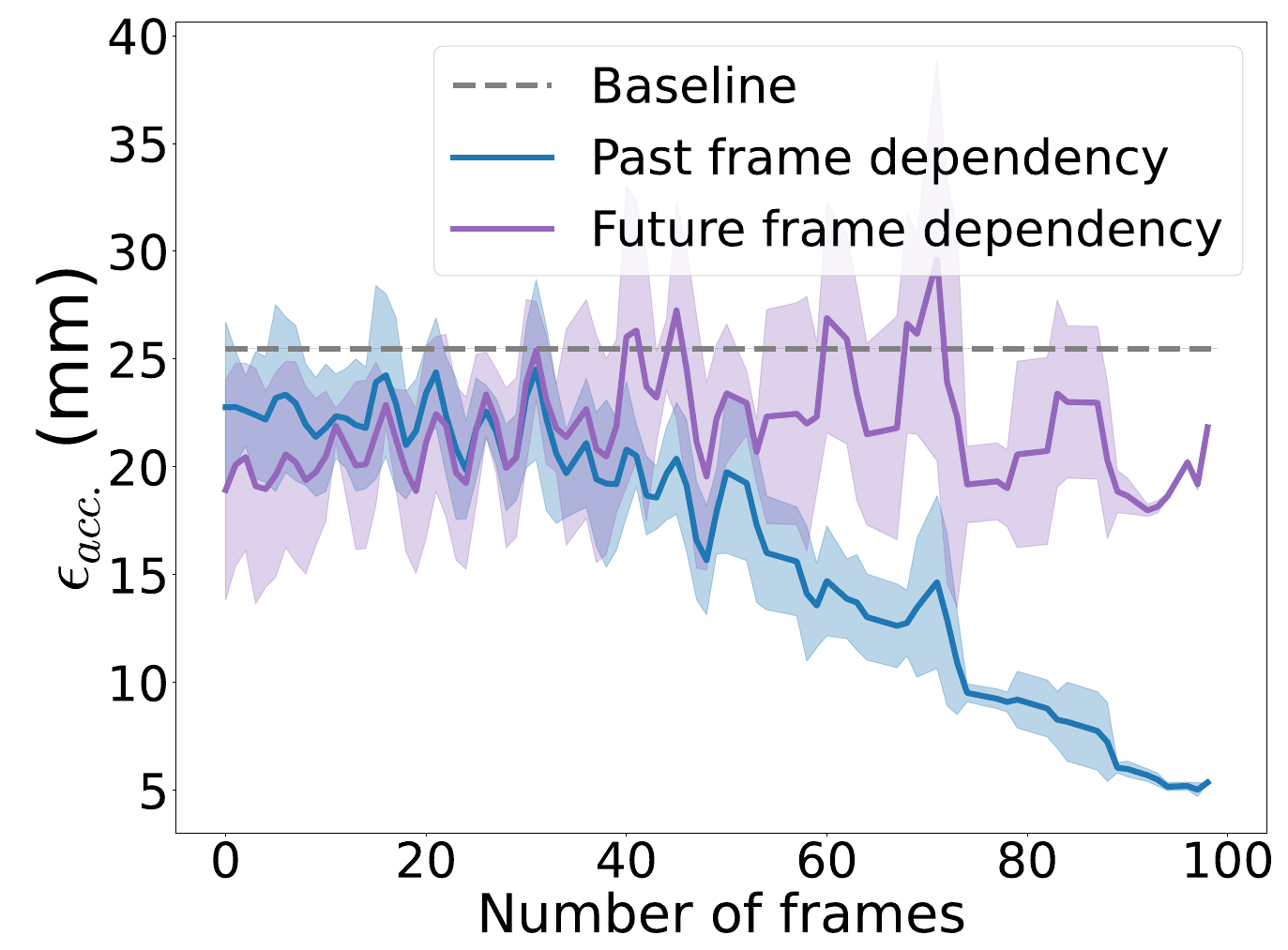}
            \caption[]%
            {{\small }}    
            \label{acc_err_reverse_CNN}
        \end{subfigure}
        
        \caption[ ]
        {$\epsilon_{acc.}$ with respect to dependency, trained using re-sampled train data with relatively constant speed (a) and a reversed frame order (b), with a CNN model. $\epsilon_{acc.}$ is shown as the mean and standard deviation over all scans in the test set, computed using models with $M=20, 49, 75, 100$.}
        \label{acc_err_sample_speed_reverse_CNN}
\end{figure}

\begin{table}[h!]
\begin{center}
\tiny
\caption{Mean and standard deviation of best performance of four metrics, among all sampled tasks, with regards to various $M$ by using CNN-based model, trained on all train data. Note: $\epsilon_{dice}$ is computed on the perpendicular scans as an example.}\label{cnn_all_train_best}%
\begin{tabular*}{\linewidth}{@{\extracolsep{\fill}}lcccc@{\extracolsep{\fill}}}
\toprule
$M$ & $\epsilon_{frame}$& $\epsilon_{acc.}$   & $\epsilon_{dice}$ &  $\epsilon_{drift}$ \\
\midrule
 2  & $0.53\pm0.46$  &$22.75\pm17.51$ &   $0.50\pm0.29$ & $29.59\pm19.53$ \\ 
   10 &  $0.46\pm0.56$   &$19.28\pm13.25$ &   $0.54\pm0.30$ & $26.55\pm13.29$ \\  
   20 &  $0.39\pm0.36$   &$16.59\pm11.45$ &   $0.58\pm0.27$ & $22.92\pm11.56$ \\  
   30 &  $0.38\pm0.35$   &$19.37\pm12.86$ &   $0.60\pm0.25$ & $27.70\pm15.70$ \\  
   40  & $0.33\pm0.35$   &$16.50\pm9.21$ &   $0.63\pm0.26$ & $23.83\pm14.95$ \\  
   49  & $0.32\pm0.22$  &$18.69\pm10.44$ &   $0.56\pm0.27$ & $28.62\pm17.62$ \\  
   60 &  $0.25\pm0.10$   &$14.08\pm8.37$ &   $0.64\pm0.26$ & $22.78\pm16.73$ \\  
   75 &  $0.24\pm0.09$   &$11.12\pm6.60$ &   $0.75\pm0.23$ & $18.20\pm14.17$ \\  
  100 &  $0.19\pm0.08$   &$4.01\pm4.01$ &   $0.77\pm0.17$ & $7.24\pm8.33$ \\

\bottomrule
\end{tabular*}
\end{center}
\end{table}

\begin{figure}[h!]
        \centering
        \begin{subfigure}[t]{0.315\linewidth}
            \centering
            
            \includegraphics[width=\linewidth]{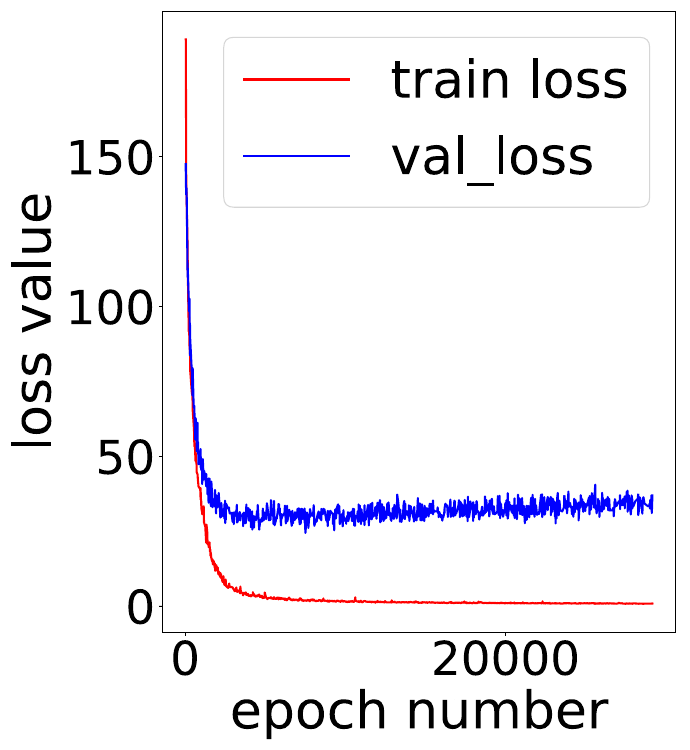}
            \caption[Network2]%
            {{\small}}    
            \label{loss_figure_all_train_data}
        \end{subfigure}
        \hfill
        \begin{subfigure}[t]{0.325\linewidth}
            \centering
            \includegraphics[width=\linewidth]{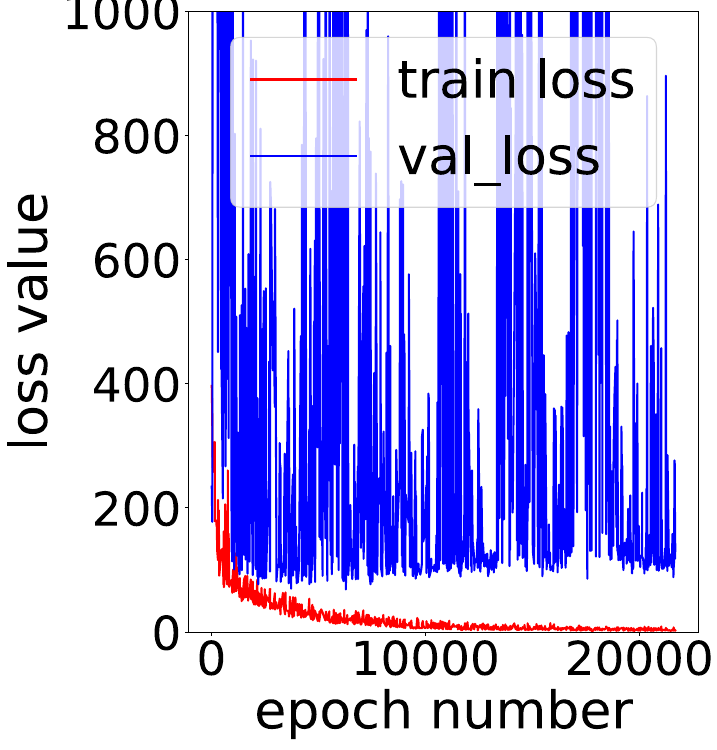}
            \caption[Network2]%
            {{\small}}    
            \label{loss_figure_linear}
        \end{subfigure}
        \hfill
        \begin{subfigure}[t]{0.335\linewidth}  
            \centering 
            \includegraphics[width=\linewidth]{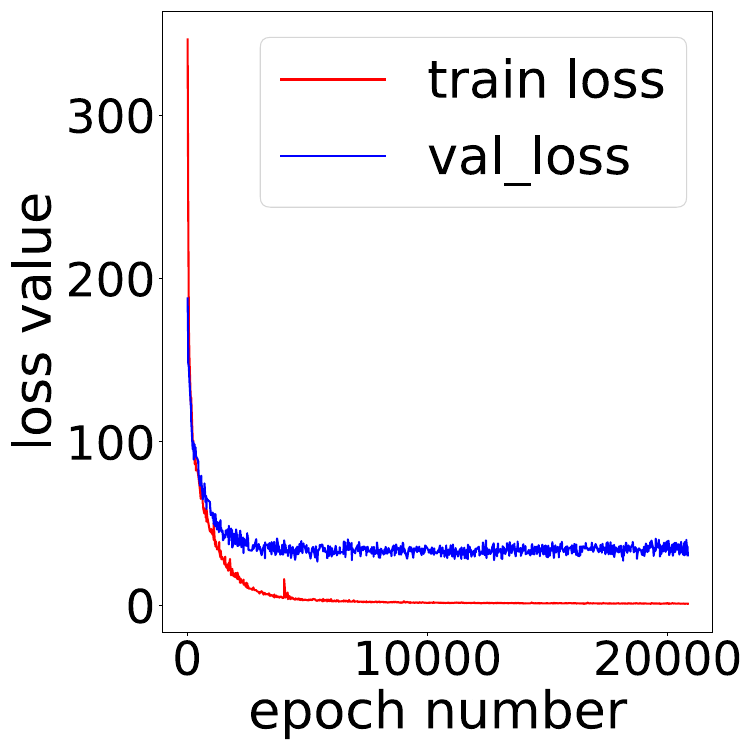}
            \caption[]%
            {{\small}}    
            \label{loss_figure_reduction_25}
        \end{subfigure}
        
        \caption[]
        {Train and validation loss of models trained with all data in train set (a), straight data in train set (b), and 25\% subject reduction of the train set (c).} 
        \label{validation_loss}
\end{figure}

\subsection{Ablation Study with Reduced Dependency Factors} \label{sec:results_varaince}

The reconstruction performance for different models trained with various variance-reduced training sets (Section \ref{sec:va_reduced}) is shown in Fig. \ref{acc_frame_pf_ff}, evaluated using $\epsilon_{acc.}$ and $\epsilon_{frame}$, with increasing past frames. In practice, the reduction in either anatomical or protocol dependencies generally led to expected poorer performance. For instance, the model trained with only `straight' scans yielded highest reconstruction errors in both $\epsilon_{acc.}$ and $\epsilon_{frame}$, worse than baseline model ($M=2$) even when the past frame number is high. This suggested that the improved reconstruction accuracy contributed by the long-term dependency (e.g., seen with more than 70 past frames), was considerably reduced by mismatched protocol variances between train and test sets. To a lesser extent, removing 75\% training subjects resulted in similar performance reduction. The other variance reduction models (`c-shape and s-shape', 75\% training subjects or 75\% scan length) yielded much less substantial performance losses, suggesting the current levels of anatomy or protocol variance still include long-term dependency and benefit the reconstruction. The same conclusion can be drawn when only perpendicular or parallel scans are sampled for testing (as shown in Fig. \ref{acc_frame_pf_ff_para_ver}).

\begin{figure*}[h!]
        \centering

        \begin{subfigure}[b]{0.49\linewidth}   
            \centering 
            \includegraphics[width=0.7\linewidth]{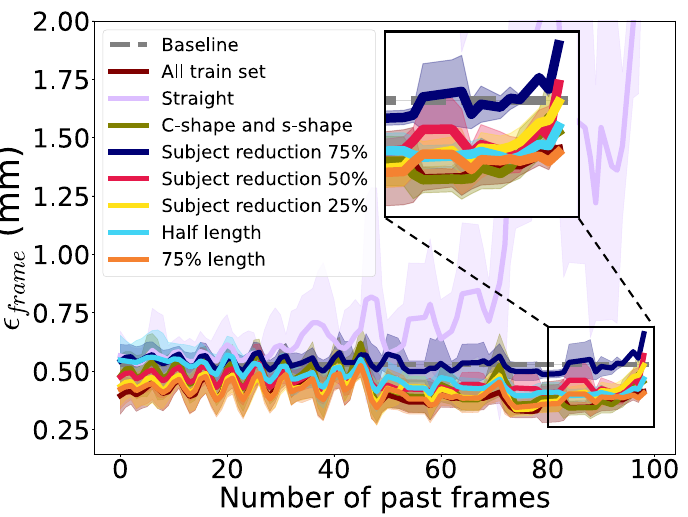}
            \caption[]%
            {{\small}}    
            \label{frame_acc_pf}
        \end{subfigure}
        \hfill
        \begin{subfigure}[b]{0.49\linewidth}   
            \centering 
            \includegraphics[width=0.7\linewidth]{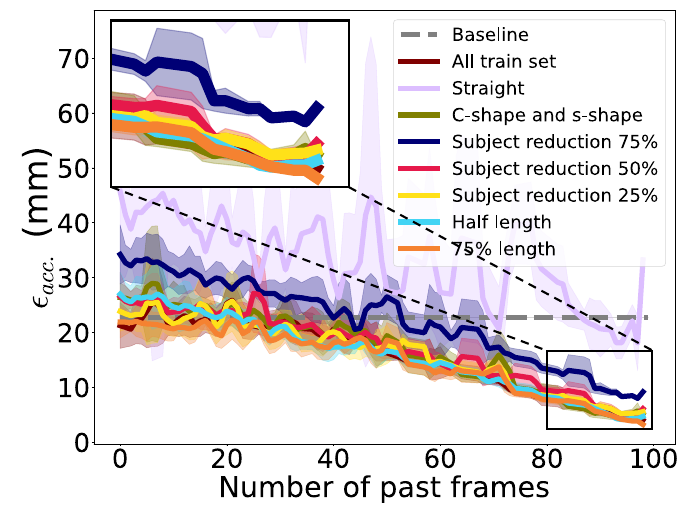}
            \caption[]%
            {{\small}}    
            \label{acc_err_pf}
        \end{subfigure}

        \caption[ $\epsilon_{frame}$. and $\epsilon_{acc.}$ with regards to past dependency.]
        {The reconstruction performance with regards to the past dependency. The performance is shown as the means and standard deviations of $\epsilon_{frame}$ and $\epsilon_{acc.}$ over all test scans, from models with $M=20, 49, 75, 100$. All models trained with different variance-reduced data are tested on the same original test set.}
        \label{acc_frame_pf_ff}
\end{figure*}

\begin{figure*}[h!]
        \centering

        \begin{subfigure}[b]{0.49\linewidth}   
            \centering 
            \includegraphics[width=0.67\linewidth]{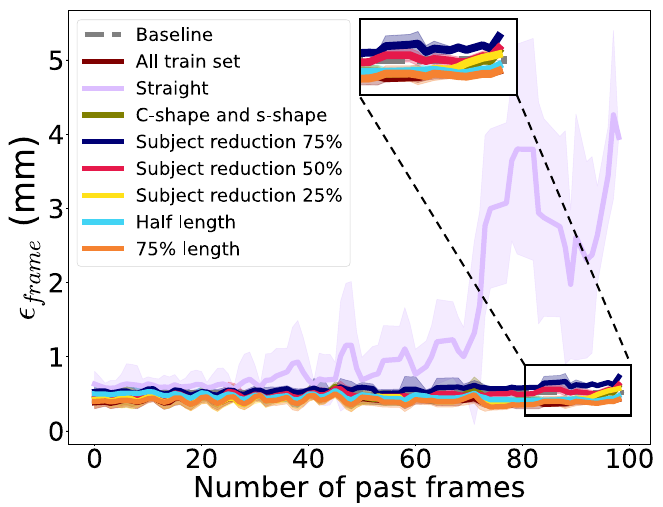}
           \caption[]%
            {{\small }} 
            \label{Li9_1}
        \end{subfigure}
        \hfill
        \begin{subfigure}[b]{0.49\linewidth}   
            \centering 
            \includegraphics[width=0.7\linewidth]{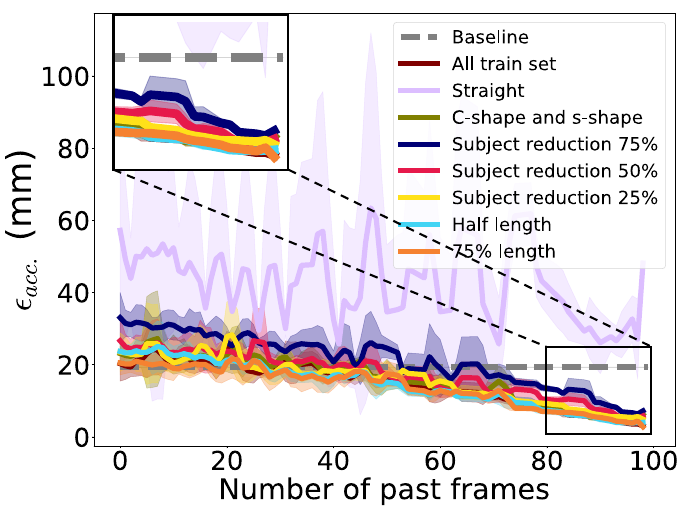}
            \caption[]%
            {{\small }}  
            \label{Li9_2}
        \end{subfigure}

        \vskip\baselineskip

        \begin{subfigure}[b]{0.49\linewidth}
            \centering
            \includegraphics[width=0.7\linewidth]{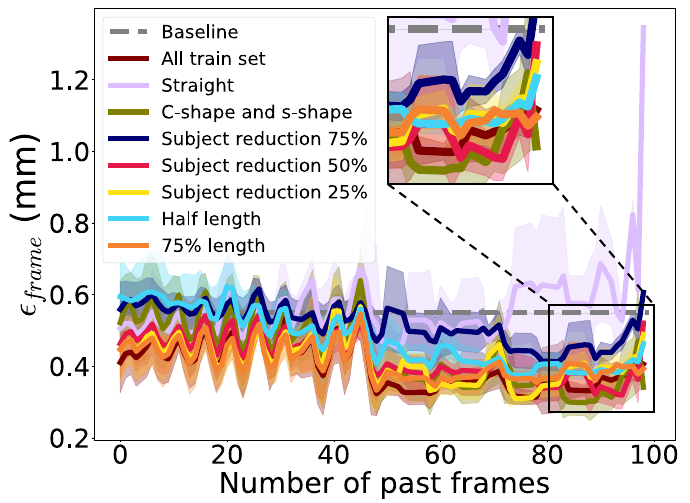}
           \caption[]%
            {{\small }} 
            \label{Li9_3}
        \end{subfigure}
        \hfill
        \begin{subfigure}[b]{0.49\linewidth}  
            \centering 
            \includegraphics[width=0.7\linewidth]{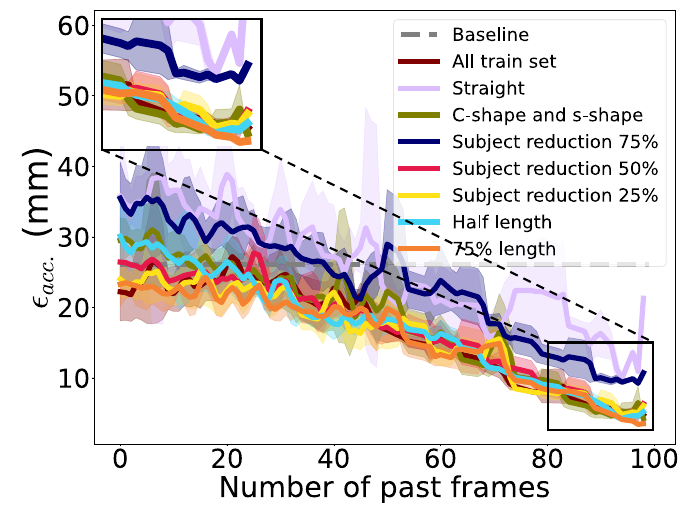}
           \caption[]%
            {{\small }}     
            \label{Li9_4}
        \end{subfigure}
        \caption[ The reconstruction performance with regards to past- and future- long-term dependency.]
        {The reconstruction performance with regards to the past long-term dependency. The performance is shown as the means and standard deviations of $\epsilon_{frame}$ and $\epsilon_{acc.}$, from models with $M=20, 49, 75, 100$, trained with different variance-reduced data, tested on parallel or perpendicular scans in the original test set: (a) Performance of $\epsilon_{frame}$ on parallel scans. (b) Performance of $\epsilon_{acc.}$ on parallel scans. (c) Performance of $\epsilon_{frame}$ on perpendicular scans. (d) Performance of $\epsilon_{acc.}$ on perpendicular scans.}

        \label{acc_frame_pf_ff_para_ver}
    \end{figure*}

\section{Discussion}\label{sec12}
Many previous studies have focused on improving the networks and their training strategies \cite{xie2021image,ning2022spatial}, including those with prior knowledge \cite{luo2021self} and additional sensors, such as IMU \cite{luo2022deep}. These developments are not necessarily specific to utilising long-term dependency, and may be applied in addition to the proposed input encoding and multi-transformation prediction to further advance the performance. The presented work adopted established CNNs and RNNs for providing the uncomplicated results to quantify the advantageous long-term dependency, as the first step towards maximising its utility. 

These results suggested that 1) Both the hypothesized anatomical and protocol dependency are likely to be factors for long-term dependency-improved reconstruction, shown in Section \ref{sec:results_varaince}; 2) Between the two, the protocol dependency is more likely to be the predominant source for the benefits from including long-term dependency; and 3) The interesting difference consistently observed between the past and future frames, in contributing to reconstruction accuracy, remains unexplained and a subject of investigation in our ongoing study.
It is important to emphasize the application-specific nature of the second conclusion.
The forearm dataset used in this paper may be considered specific in a number of aspects, including anatomical content richness, compared with other anatomical targets. As a result, conclusions drawn in this study may need further evaluation, when a different data set is considered. The proposed methodology however shall still be applicable and fit-for-purpose for different clinical indications. Future work should thus aim to access tracked ultrasound data from different clinical usages.
Furthermore, 3D visualisation in addition to its geometric reconstruction of US volume can be crucial for many applications, which should be investigated further for its clinical applicability.
It is worth noting that the influence of scanning speed, image quality, and overlap between adjacent frames remains open research questions for freehand US reconstruction, and may affect the robustness and generalisation of the tested reconstruction method, in turn may be specific to our conclusion on long-term dependency. In addition, definitions for these factors also remain an open research question and multiple potential solutions have been proposed, such as protocol-adaptation\cite{chen2020cross} and application-specific quality-control\cite{saeed2022image}. The generalisability of our conclusion should be interpreted with respect to individual applications and reconstruction algorithms, although the methodology of factorising the long-term dependency may nevertheless be useful.
Besides, quantifying the difference between the anatomy and protocol dependency can also be application-dependent, with varying practical costs for changing their complexity. Further experimental results for these applications are mandatory.
What is more, US images acquired using different ultrasound scanner/probe, by different researchers, may be different due to the specific parameter configurations and the intra- and inter-variability of the acquired US images. Therefore, acquiring various types of dataset by using various kinds of ultrasound scanner/probe, by a number of researchers, is our future research focus, which can be used to test the generalization of the conclusion.

\section{Conclusion}\label{sec13}
This work proposed a new parametric dependency based on frame encoding and multi-tasking transformation, to quantify dependency factors originated from anatomical and protocol characteristics. The experiments showed that long-term dependency based on either recurrent or feed-forward models can significantly improve reconstruction, and the improvement was dependent on the frame-to-transformation distance and transformation intervals. It was also found that the scanning protocol and, to a lesser degree, the anatomical content are both important in utilising the long-term dependency. As the proposed approach is based on the multi-task learning framework, negative transfer among tasks could lead to inferior performance. The ongoing work investigates methods such as task grouping \cite{fifty2021efficiently} to further improve the main task and for potentially utilising multiple main tasks.

\section*{Acknowledgment}
This work was supported by the EPSRC [EP/T029404/1], a Royal Academy of Engineering / Medtronic Research Chair [RCSRF1819\textbackslash7\textbackslash734] (TV), Wellcome/EPSRC Centre for Interventional and Surgical Sciences [203145Z/16/Z], and the International Alliance for Cancer Early Detection, an alliance between Cancer Research UK [C28070/A30912; C73666/A31378], Canary Center at Stanford University, the University of Cambridge, OHSU Knight Cancer Institute, University College London and the University of Manchester. TV is co-founder and shareholder of Hypervision Surgical. Qi Li was supported by the University College London Overseas and Graduate Research Scholarships. 
For the purpose of Open Access, the authors have applied a CC BY public copyright licence to any Author Accepted Manuscript version arising from this submission.


\bibliographystyle{IEEEbib}
\bibliography{ref}




\clearpage
\renewcommand{\theequation}{S\arabic{equation}}
\renewcommand{\thetable}{S\arabic{table}}
\renewcommand{\thefigure}{S\arabic{figure}}
\setcounter{equation}{0}
\setcounter{table}{0}
\setcounter{figure}{0}

\section*{Supplementary Materials}


\begin{figure*}[h!]
\renewcommand{\thefigure}{S-\arabic{figure}}
        \centering 
        \begin{subfigure}[b]{0.42\textwidth}   
            \centering 
            \includegraphics[width=0.98\textwidth]{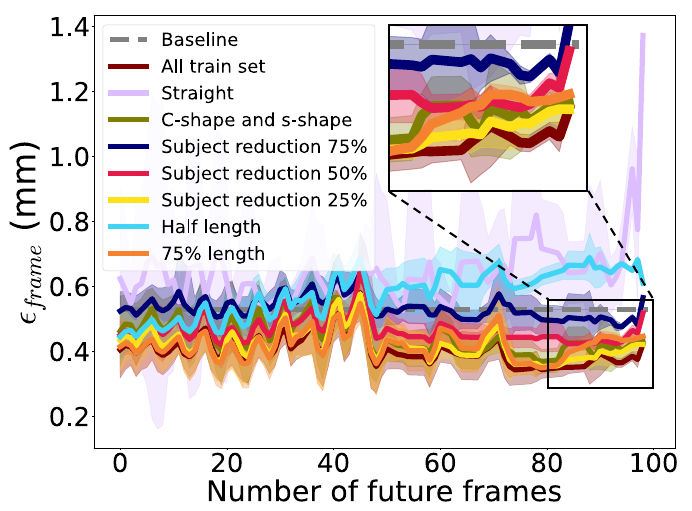}
            \caption[]%
            {{\small}}    
            \label{Li_S1_1}
        \end{subfigure}
        \hfill
        \begin{subfigure}[b]{0.42\textwidth}   
            \centering 
            \includegraphics[width=0.98\textwidth]{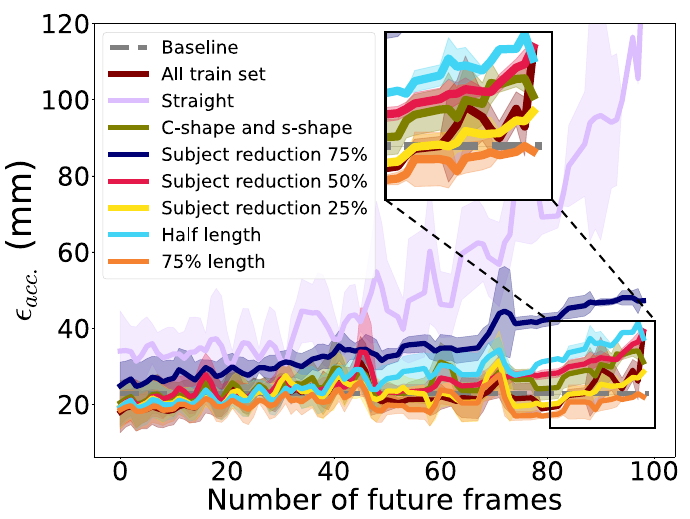}
            \caption[]%
            {{\small}}    
            \label{Li_S1_2}
        \end{subfigure} 
        \caption[ The reconstruction performance with regards to future long-term dependency.]
        {The reconstruction performance with regards to future long-term dependency. The performance is shown as the mean and standard deviation of $\epsilon_{frame}$ and $\epsilon_{acc.}$ over all scans in the test set, from models with $M=20, 49, 75, 100$. All models trained with different variance-reduced data are tested on the same original test set. (a) Performance of $\epsilon_{frame}$ with regards to the number of future frames. (b) Performance of $\epsilon_{acc.}$ with regards to the number of future frames.}
        \label{acc_frame_pf_ff-s}
    \end{figure*}
\begin{figure*}[h!]
\renewcommand{\thefigure}{S-\arabic{figure}}
        \centering
        \begin{subfigure}[b]{0.42\textwidth}   
            \centering 
            \includegraphics[width=0.98\textwidth]{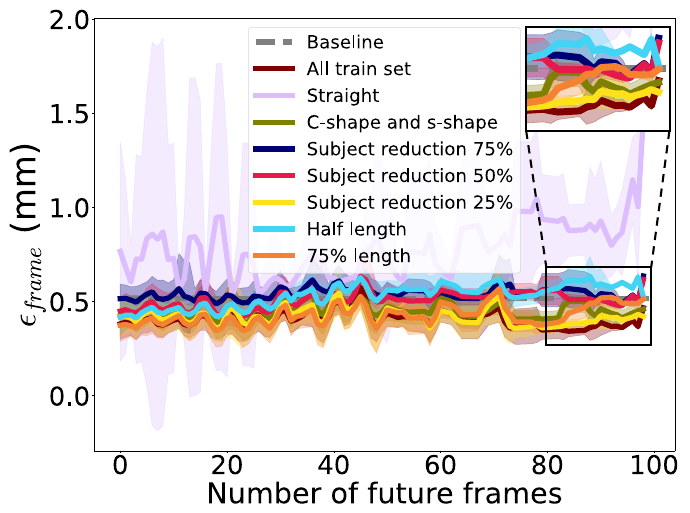}
           \caption[]%
            {{\small }} 
            \label{frame_acc_pf-s}
        \end{subfigure}
        \hfill
        \begin{subfigure}[b]{0.42\textwidth}   
            \centering 
            \includegraphics[width=0.98\textwidth]{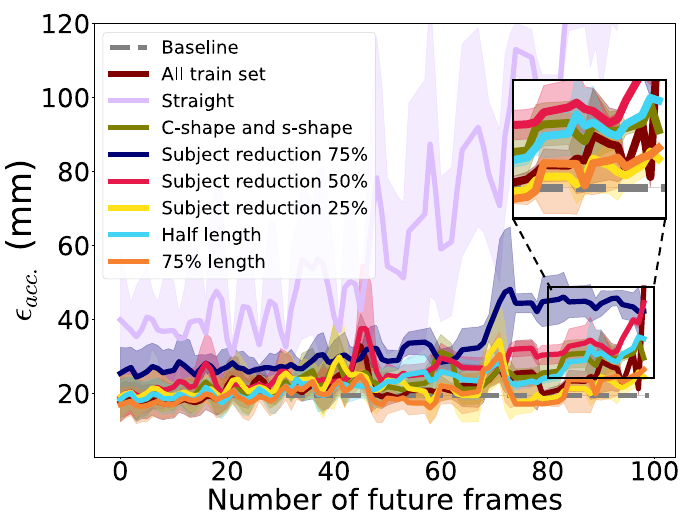}
            \caption[]%
            {{\small }}  
            \label{frame_acc_ff-s}
        \end{subfigure}        
        \vskip\baselineskip        
        \begin{subfigure}[b]{0.42\textwidth}
            \centering
            \includegraphics[width=0.98\textwidth]{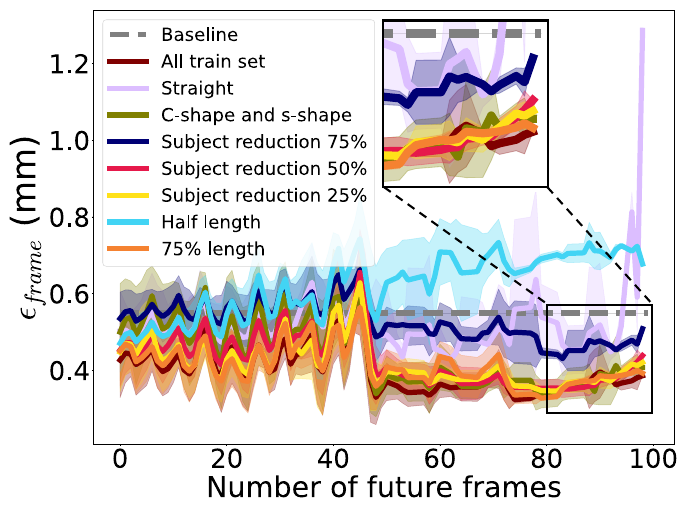}
           \caption[]%
            {{\small }} 
            \label{acc_err_pf-s}
        \end{subfigure}
        \hfill
        \begin{subfigure}[b]{0.42\textwidth}  
            \centering 
            \includegraphics[width=0.98\textwidth]{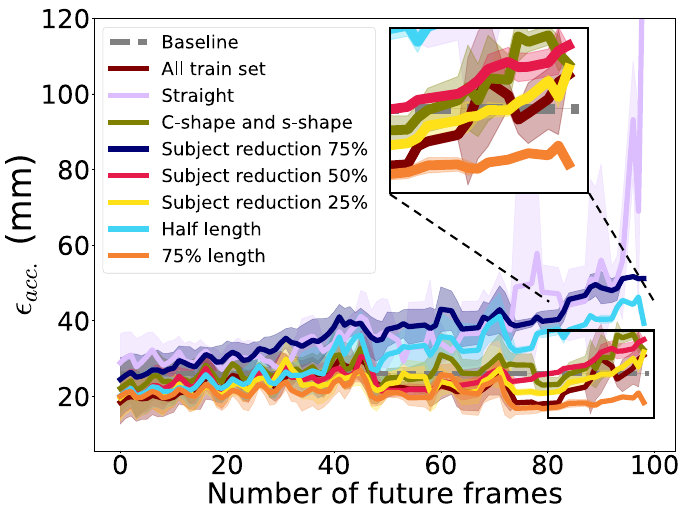}
           \caption[]%
            {{\small }}     
            \label{acc_err_ff-s}
        \end{subfigure}
        \caption[The reconstruction performance with regards to past- and future- long-term dependency.]
        {The reconstruction performance with regards to the future long-term dependency. The performance is shown as the mean and standard deviation of $\epsilon_{frame}$ and $\epsilon_{acc.}$, from models with $M=20, 49, 75, 100$, trained with different variance-reduced data, tested on parallel or perpendicular scans in the original test set: (a) Performance of $\epsilon_{frame}$ on parallel scans. (b) Performance of $\epsilon_{acc.}$ on parallel scans. (c) Performance of $\epsilon_{frame}$ on perpendicular scans. (d) Performance of $\epsilon_{acc.}$ on perpendicular scans.}
        \label{acc_frame_ff_para_ver-s}
    \end{figure*}
    
\begin{table}[H]
\renewcommand{\thetable}{S-\Roman{table}}
\begin{center}
\tiny
\caption{Mean and standard deviation of average performance of four metrics, over all sampled tasks, with regards to various $M$ by using CNN-based model, trained on all train data.}\label{cnn_all_train_average-s}%

\begin{tabular*}{\linewidth}{@{\extracolsep{\fill}}lcccc@{\extracolsep{\fill}}}
\toprule
$M$ & $\epsilon_{frame}$& $\epsilon_{acc.}$   & $\epsilon_{dice}$ &  $\epsilon_{drift}$\\
\midrule
    2  & $0.53\pm0.00$   &$22.75\pm0.00$ &  $0.50\pm0.00$ &  $29.59\pm0.00$ \\
   10  & $0.51\pm0.03$   &$20.29\pm1.48$ &  $0.46\pm0.07$ &  $28.15\pm1.92$ \\ 
   20  & $0.45\pm0.04$   &$18.91\pm2.10$ &  $0.46\pm0.08$ &  $26.79\pm3.33$ \\
   30  & $0.47\pm0.06$   &$21.37\pm1.32$ &  $0.41\pm0.10$ &  $30.56\pm1.96$ \\
   40  & $0.46\pm0.06$   &$21.13\pm2.20$ &  $0.43\pm0.10$ &  $30.80\pm3.00$ \\
   49 &  $0.46\pm0.08$   &$23.89\pm3.88$ &  $0.37\pm0.10$ &  $35.04\pm5.13$ \\ 
   60  & $0.39\pm0.09$  &$21.16\pm5.20$ &  $0.41\pm0.12$ &  $32.48\pm7.57$ \\ 
   75  & $0.37\pm0.07$   &$21.29\pm4.91$ &  $0.34\pm0.16$ &  $29.17\pm6.43$ \\ 
  100 &  $0.32\pm0.06$   &$16.35\pm6.33$ &  $0.23\pm0.15$ &  $20.26\pm7.19$ \\

\bottomrule
\end{tabular*}
\end{center}
\end{table}
\noindent

\begin{table}[H]
\renewcommand{\thetable}{S-\Roman{table}}
\begin{center}
\tiny
\caption{Mean and standard deviation of best performance of four metrics, over all sampled tasks, with regards to various $M$ by using RNN-based model, trained on all train data.}\label{rnn_all_train_best-s}%
\begin{tabular*}{\linewidth}{@{\extracolsep{\fill}}lcccc@{\extracolsep{\fill}}}
\toprule
$M$ & $\epsilon_{frame}$& $\epsilon_{acc.}$   & $\epsilon_{dice}$ & $\epsilon_{drift}$\\ 
\midrule

    2  &  $0.57\pm0.44$   &$26.33\pm15.99$ &   $0.41\pm0.33$ & $34.54\pm18.10$ \\  
   10 &   $0.37\pm0.30$   &$17.29\pm11.13$ &   $0.62\pm0.24$ & $24.35\pm13.57$ \\ 
   20 &   $0.37\pm0.29$  &$16.29\pm9.82$ &   $0.61\pm0.26$ & $24.98\pm13.11$ \\  
   30 &   $0.37\pm0.31$   &$19.36\pm11.53$ &   $0.59\pm0.26$ & $28.53\pm15.85$ \\  
   40 &   $0.30\pm0.15$   &$17.07\pm8.97$ &   $0.61\pm0.26$ & $26.96\pm15.65$ \\  
   49 &   $0.27\pm0.16$   &$15.32\pm7.89$ &   $0.65\pm0.22$ & $24.21\pm15.84$ \\  
   60 &   $0.26\pm0.10$  &$14.53\pm7.64$ &   $0.66\pm0.21$ & $22.58\pm14.15$ \\  
   75 &    $0.23\pm0.10$   &$10.64\pm6.09$ &   $0.77\pm0.12$ & $17.09\pm11.24$ \\  
  100  &  $0.20\pm0.07$   &$4.27\pm3.66$ &   $0.73\pm0.23$ & $6.97\pm6.79$\\

\bottomrule
\end{tabular*}
\end{center}
\end{table}
\noindent

\begin{table}[H]
\renewcommand{\thetable}{S-\Roman{table}}
\begin{center}
\tiny
\caption{Mean and standard deviation of average performance of four metrics, over all sampled tasks, with regards to various $M$ by using RNN-based model, trained on all train data.}\label{rnn_all_train_average-s}%
\begin{tabular*}{\linewidth}{@{\extracolsep{\fill}}lcccc@{\extracolsep{\fill}}}
\toprule
$M$ & $\epsilon_{frame}$& $\epsilon_{acc.}$   & $\epsilon_{dice}$ &  $\epsilon_{drift}$\\
\midrule

 2  &  $0.57\pm0.00$ &  $26.33\pm0.00$ &  $0.41\pm0.00$ &  $34.54\pm0.00$ \\
 10 &   $0.42\pm0.03$ &  $18.53\pm0.82$ &  $0.50\pm0.09$ &  $25.74\pm1.25$ \\ 
 20 &   $0.45\pm0.05$   &$18.14\pm1.07$ &  $0.48\pm0.08$ &  $27.36\pm1.51$ \\ 
 30  &  $0.48\pm0.05$   &$21.44\pm1.11$ &  $0.43\pm0.10$ &  $31.24\pm1.74$ \\ 
 40 &   $0.43\pm0.08$   &$21.24\pm3.33$ &  $0.42\pm0.11$ &  $31.91\pm4.75$ \\ 
 49 &   $0.45\pm0.11$   &$21.38\pm5.31$ &  $0.43\pm0.11$ &  $32.59\pm7.79$ \\
 60  &  $0.39\pm0.08$   &$20.84\pm3.78$ &  $0.41\pm0.13$ &  $31.23\pm5.61$ \\ 
 75 &   $0.38\pm0.09$   &$20.75\pm4.97$ &  $0.40\pm0.13$ &  $30.44\pm7.03$ \\ 
100  &  $0.34\pm0.09$   &$16.80\pm6.78$ &  $0.22\pm0.13$ &  $20.87\pm7.80$ \\ 
\bottomrule
\end{tabular*}
\end{center}
\end{table}
\noindent

\begin{table}[H]
\renewcommand{\thetable}{S-\Roman{table}}
\begin{center}
\tiny
\caption{Mean and standard deviation of best performance of four metrics, among all sampled tasks, with regards to various $M$ by using CNN-based model, trained on straight train data.}\label{cnn_straight_best-s}%
\begin{tabular*}{\linewidth}{@{\extracolsep{\fill}}lcccc@{\extracolsep{\fill}}}
\toprule
$M$ & $\epsilon_{frame}$& $\epsilon_{acc.}$   & $\epsilon_{dice}$ &  $\epsilon_{drift}$ \\
\midrule

   20   & $0.51\pm0.31$ &    $31.84\pm17.16$ &  $0.45\pm0.29$ &   $39.88\pm19.33$   \\
   49   & $0.43\pm0.26$ &    $23.83\pm14.05$ &  $0.57\pm0.25$ &   $35.22\pm25.10$  \\ 
   75   & $0.40\pm0.24$ &    $13.06\pm7.09$ &  $0.66\pm0.22$ &   $22.77\pm17.92$   \\
  100   & $0.48\pm0.25$ &    $13.00\pm23.79$ &  $0.64\pm0.26$ &   $22.30\pm41.10$ \\
\bottomrule
\end{tabular*}
\end{center}
\end{table}
\noindent

\begin{table}[H]
\renewcommand{\thetable}{S-\Roman{table}}
\begin{center}
\tiny
\caption{Mean and standard deviation of average performance of four metrics, over all sampled tasks, with regards to various $M$ by using CNN-based model, trained on straight train data.}\label{cnn_straight_average-s}%
\begin{tabular*}{\linewidth}{@{\extracolsep{\fill}}lcccc@{\extracolsep{\fill}}}
\toprule
$M$ & $\epsilon_{frame}$& $\epsilon_{acc.}$   & $\epsilon_{dice}$ &  $\epsilon_{drift}$ \\
\midrule

  20   & $0.57\pm0.04$ &    $39.85\pm2.03$ &  $0.25\pm0.10$ &   $48.20\pm3.42$    \\
   49   & $0.55\pm0.07$ &    $31.93\pm4.02$ &  $0.29\pm0.11$ &   $42.33\pm4.86$    \\
   75   & $0.48\pm0.07$ &    $30.28\pm9.17$ &  $0.31\pm0.16$ &   $38.83\pm10.42$    \\
  100   & $0.92\pm0.52$ &    $58.46\pm22.81$ &  $0.13\pm0.10$ &   $59.04\pm24.95$   \\ 

\bottomrule
\end{tabular*}
\end{center}
\end{table}
\noindent

\begin{table}[H]
\renewcommand{\thetable}{S-\Roman{table}}
\begin{center}
\tiny
\caption{Mean and standard deviation of best performance of four metrics, among all sampled tasks, with regards to various $M$ by using CNN-based model, trained on c-shape and s-shape train data.}\label{cnn_c_s_best-s}%
\begin{tabular*}{\linewidth}{@{\extracolsep{\fill}}lcccc@{\extracolsep{\fill}}}
\toprule
$M$ & $\epsilon_{frame}$& $\epsilon_{acc.}$   & $\epsilon_{dice}$ &  $\epsilon_{drift}$ \\
\midrule
   20   & $0.48\pm0.65$ &    $23.17\pm19.62$ &  $0.51\pm0.32$ &   $30.29\pm18.02$   \\
   49   & $0.40\pm0.40$ &    $17.62\pm8.39$ &  $0.60\pm0.28$ &   $28.17\pm20.22$   \\
   75   & $0.25\pm0.13$ &    $10.04\pm6.78$ &  $0.78\pm0.14$ &   $17.10\pm13.16$  \\ 
  100   & $0.24\pm0.13$ &    $4.00\pm2.72$ &  $0.80\pm0.13$ &   $6.74\pm7.19$   \\

\bottomrule
\end{tabular*}
\end{center}
\end{table}
\noindent

\begin{table}[H]
\renewcommand{\thetable}{S-\Roman{table}}
\begin{center}
\tiny
\caption{Mean and standard deviation of average performance of four metrics, over all sampled tasks, with regards to various $M$ by using CNN-based model, trained on c-shape and s-shape train data.}\label{cnn_c_s_average-s}%
\begin{tabular*}{\linewidth}{@{\extracolsep{\fill}}lcccc@{\extracolsep{\fill}}}
\toprule
$M$ & $\epsilon_{frame}$& $\epsilon_{acc.}$   & $\epsilon_{dice}$ &  $\epsilon_{drift}$ \\
\midrule

   20   & $0.55\pm0.04$ &    $25.02\pm1.84$ &  $0.38\pm0.08$ &   $33.06\pm3.10$   \\
   49   & $0.54\pm0.08$ &    $24.71\pm4.92$ &  $0.35\pm0.10$ &   $35.96\pm6.67$   \\ 
   75   & $0.39\pm0.08$ &    $22.68\pm4.86$ &  $0.34\pm0.14$ &   $32.30\pm6.38$   \\ 
  100   & $0.33\pm0.07$ &    $17.38\pm7.54$ &  $0.25\pm0.14$ &   $21.34\pm8.36$    \\

\bottomrule
\end{tabular*}
\end{center}
\end{table}
\noindent

\begin{table}[H]
\renewcommand{\thetable}{S-\Roman{table}}
\begin{center}
\tiny
\caption{Mean and standard deviation of best performance of four metrics, among all sampled tasks, with regards to various $M$ by using CNN-based model, trained on train data with subjects reduction rate of 25\%.}\label{cnn_reduce25_best-s}%
\begin{tabular*}{\linewidth}{@{\extracolsep{\fill}}lcccc@{\extracolsep{\fill}}}
\toprule
$M$ & $\epsilon_{frame}$& $\epsilon_{acc.}$   & $\epsilon_{dice}$ &  $\epsilon_{drift}$ \\
\midrule

   20   & $0.44\pm0.40$ &    $19.64\pm9.04$ &  $0.53\pm0.30$ &   $27.42\pm10.79$ \\  
   49   & $0.32\pm0.32$ &    $15.27\pm7.48$ &  $0.63\pm0.28$ &   $24.88\pm15.80$   \\
   75   & $0.24\pm0.12$ &    $10.35\pm6.27$ &  $0.73\pm0.13$ &   $17.31\pm11.62$  \\ 
  100   & $0.20\pm0.07$ &    $4.49\pm4.21$ &  $0.76\pm0.20$ &   $8.40\pm9.08$   \\

\bottomrule
\end{tabular*}
\end{center}
\end{table}
\noindent

\begin{table}[H]
\renewcommand{\thetable}{S-\Roman{table}}
\begin{center}
\tiny
\caption{Mean and standard deviation of average performance of four metrics, over all sampled tasks, with regards to various $M$ by using CNN-based model, trained on train data with subjects reduction rate of 25\%.}\label{cnn_reduce25_average-s}%
\begin{tabular*}{\linewidth}{@{\extracolsep{\fill}}lcccc@{\extracolsep{\fill}}}
\toprule
$M$ & $\epsilon_{frame}$& $\epsilon_{acc.}$   & $\epsilon_{dice}$ &  $\epsilon_{drift}$ \\
\midrule
   20   & $0.50\pm0.04$ &    $22.45\pm2.64$ &  $0.39\pm0.08$ &   $31.87\pm3.94$ \\   
   49   & $0.46\pm0.09$ &    $22.89\pm4.64$ &  $0.40\pm0.10$ &   $34.06\pm6.47$ \\   
   75   & $0.40\pm0.08$ &    $22.37\pm5.34$ &  $0.34\pm0.15$ &   $31.87\pm7.30$ \\  
  100   & $0.32\pm0.06$ &    $15.49\pm5.15$ &  $0.24\pm0.15$ &   $19.74\pm5.96$ \\ 
 
\bottomrule
\end{tabular*}
\end{center}
\end{table}
\noindent

\begin{table}[H]
\renewcommand{\thetable}{S-\Roman{table}}
\begin{center}
\tiny
\caption{Mean and standard deviation of best performance of four metrics, among all sampled tasks, with regards to various $M$ by using CNN-based model, trained on train data with subjects reduction rate of 50\%.}\label{cnn_reduce50_best-s}%
\begin{tabular*}{\linewidth}{@{\extracolsep{\fill}}lcccc@{\extracolsep{\fill}}}
\toprule
$M$ & $\epsilon_{frame}$& $\epsilon_{acc.}$   & $\epsilon_{dice}$ &  $\epsilon_{drift}$ \\
\midrule
   20   & $0.47\pm0.42$ &    $20.56\pm9.62$ &  $0.49\pm0.29$ &   $30.02\pm15.26$   \\
   49   & $0.34\pm0.19$ &    $16.07\pm9.01$ &  $0.57\pm0.22$ &   $24.28\pm17.53$  \\ 
   75   & $0.27\pm0.11$ &    $10.22\pm6.42$ &  $0.71\pm0.16$ &   $19.22\pm13.09$  \\ 
  100   & $0.29\pm0.20$ &    $4.02\pm3.82$ &  $0.83\pm0.08$ &   $7.61\pm9.38$ \\

\bottomrule
\end{tabular*}
\end{center}
\end{table}
\noindent

\begin{table}[H]
\renewcommand{\thetable}{S-\Roman{table}}
\begin{center}
\tiny
\caption{Mean and standard deviation of average performance of four metrics, over all sampled tasks, with regards to various $M$ by using CNN-based model, trained on train data with subjects reduction rate of 50\%.}\label{cnn_reduce50_average-s}%
\begin{tabular*}{\linewidth}{@{\extracolsep{\fill}}lcccc@{\extracolsep{\fill}}}
\toprule
$M$ & $\epsilon_{frame}$& $\epsilon_{acc.}$   & $\epsilon_{dice}$ &  $\epsilon_{drift}$ \\
\midrule

   20   & $0.53\pm0.05$ &    $24.46\pm4.46$ &  $0.38\pm0.07$ &   $34.87\pm6.09$ \\   
   49   & $0.50\pm0.08$ &    $24.35\pm4.91$ &  $0.34\pm0.10$ &   $35.12\pm7.16$ \\   
   75   & $0.41\pm0.07$ &    $22.13\pm4.73$ &  $0.33\pm0.15$ &   $31.47\pm5.84$ \\   
  100   & $0.40\pm0.06$ &    $19.19\pm7.27$ &  $0.24\pm0.16$ &   $22.73\pm7.80$ \\  

\bottomrule
\end{tabular*}
\end{center}
\end{table}
\noindent

\begin{table}[H]
\renewcommand{\thetable}{S-\Roman{table}}
\begin{center}
\tiny
\caption{Mean and standard deviation of best performance of four metrics, among all sampled tasks, with regards to various $M$ by using CNN-based model, trained on train data with subjects reduction rate of 75\%.}\label{cnn_reduce75_best-s}%
\begin{tabular*}{\linewidth}{@{\extracolsep{\fill}}lcccc@{\extracolsep{\fill}}}
\toprule
$M$ & $\epsilon_{frame}$& $\epsilon_{acc.}$   & $\epsilon_{dice}$ &  $\epsilon_{drift}$ \\
\midrule

   20   & $0.55\pm0.49$ &    $26.62\pm14.07$ &  $0.50\pm0.25$ &   $37.02\pm16.19$ \\  
   49   & $0.42\pm0.34$ &    $18.51\pm9.72$ &  $0.56\pm0.30$ &   $29.50\pm19.37$   \\
   75   & $0.32\pm0.21$ &    $12.01\pm7.46$ &  $0.62\pm0.28$ &   $20.93\pm13.01$ \\  
  100   & $0.41\pm0.24$ &    $7.83\pm9.08$ &  $0.75\pm0.17$ &   $13.66\pm15.94$  \\ 

\bottomrule
\end{tabular*}
\end{center}
\end{table}
\noindent

\begin{table}[H]
\renewcommand{\thetable}{S-\Roman{table}}
\begin{center}
\tiny
\caption{Mean and standard deviation of average performance of four metrics, over all sampled tasks, with regards to various $M$ by using CNN-based model, trained on train data with subjects reduction rate of 75\%.}\label{cnn_reduce75_average-s}%
\begin{tabular*}{\linewidth}{@{\extracolsep{\fill}}lcccc@{\extracolsep{\fill}}}
\toprule
$M$ & $\epsilon_{frame}$& $\epsilon_{acc.}$   & $\epsilon_{dice}$ &  $\epsilon_{drift}$\\
\midrule

   20   & $0.58\pm0.02$ &    $30.03\pm1.37$ &  $0.36\pm0.08$ &   $41.24\pm2.45$ \\  
   49   & $0.56\pm0.06$ &    $28.53\pm4.23$ &  $0.30\pm0.10$ &   $39.93\pm5.13$ \\  
   75   & $0.50\pm0.09$ &    $28.75\pm7.59$ &  $0.24\pm0.13$ &   $39.82\pm10.05$ \\   
  100   & $0.48\pm0.06$ &    $28.38\pm10.89$ &  $0.15\pm0.15$ &   $32.40\pm10.87$ \\ 

\bottomrule
\end{tabular*}
\end{center}
\end{table}
\noindent

\begin{table}[H]
\renewcommand{\thetable}{S-\Roman{table}}
\begin{center}
\tiny
\caption{Mean and standard deviation of best performance of four metrics, among all sampled tasks, with regards to various $M$ by using CNN-based model, trained on train data with half length.}\label{cnn_half_len_best-s}%
\begin{tabular*}{\linewidth}{@{\extracolsep{\fill}}lcccc@{\extracolsep{\fill}}}
\toprule
$M$ & $\epsilon_{frame}$& $\epsilon_{acc.}$   & $\epsilon_{dice}$ &  $\epsilon_{drift}$ \\
\midrule

   20   & $0.45\pm0.40$ &    $22.58\pm12.88$ &  $0.55\pm0.25$ &   $31.77\pm20.22$   \\
   49   & $0.36\pm0.31$ &    $15.72\pm7.17$ &  $0.63\pm0.21$ &   $24.83\pm15.46$   \\
   75   & $0.28\pm0.09$ &    $10.79\pm6.19$ &  $0.70\pm0.22$ &   $19.06\pm14.47$  \\ 
  100   & $0.21\pm0.08$ &    $3.80\pm3.42$ &  $0.72\pm0.27$ &   $6.54\pm6.81$   \\

\bottomrule
\end{tabular*}
\end{center}
\end{table}
\noindent

\begin{table}[H]
\renewcommand{\thetable}{S-\Roman{table}}
\begin{center}
\tiny
\caption{Mean and standard deviation of average performance of four metrics, over all sampled tasks, with regards to various $M$ by using CNN-based model, trained on train data with half length.}\label{cnn_half_len_average-s}%
\begin{tabular*}{\linewidth}{@{\extracolsep{\fill}}lcccc@{\extracolsep{\fill}}}
\toprule
$M$ & $\epsilon_{frame}$& $\epsilon_{acc.}$   & $\epsilon_{dice}$ &  $\epsilon_{drift}$ \\
\midrule

   20   & $0.53\pm0.05$ &    $23.89\pm1.32$ &  $0.40\pm0.08$ &   $34.67\pm2.29$ \\   
   49   & $0.53\pm0.08$ &    $22.84\pm3.59$ &  $0.35\pm0.11$ &   $34.70\pm5.04$ \\   
   75   & $0.51\pm0.11$ &    $23.13\pm6.55$ &  $0.28\pm0.16$ &   $33.85\pm8.66$ \\  
  100   & $0.43\pm0.12$ &    $18.11\pm8.75$ &  $0.15\pm0.15$ &   $22.78\pm10.07$ \\ 

\bottomrule
\end{tabular*}
\end{center}
\end{table}
\noindent

\begin{table}[H]
\renewcommand{\thetable}{S-\Roman{table}}
\begin{center}
\tiny
\caption{Mean and standard deviation of best performance of four metrics, among all sampled tasks, with regards to various $M$ by using CNN-based model, trained on train data with 75\% length.}\label{cnn_75_len_best-s}%
\begin{tabular*}{\linewidth}{@{\extracolsep{\fill}}lcccc@{\extracolsep{\fill}}}
\toprule
$M$ & $\epsilon_{frame}$& $\epsilon_{acc.}$   & $\epsilon_{dice}$ &  $\epsilon_{drift}$ \\
\midrule

   20   & $0.41\pm0.37$ &    $18.76\pm11.11$ &  $0.53\pm0.25$ &   $27.33\pm13.54$  \\ 
   49   & $0.31\pm0.27$ &    $17.15\pm9.87$ &  $0.58\pm0.28$ &   $26.48\pm19.58$  \\ 
   75   & $0.23\pm0.10$ &    $12.16\pm8.40$ &  $0.73\pm0.15$ &   $20.24\pm15.60$  \\ 
  100   & $0.19\pm0.06$ &    $3.19\pm2.36$ &  $0.78\pm0.13$ &   $6.46\pm5.39$   \\

\bottomrule
\end{tabular*}
\end{center}
\end{table}
\noindent

\begin{table}[H]
\renewcommand{\thetable}{S-\Roman{table}}
\begin{center}
\tiny
\caption{Mean and standard deviation of average performance of four metrics, over all sampled tasks, with regards to various $M$ by using CNN-based model, trained on train data with 75\% length.}\label{cnn_75_len_average-s}%
\begin{tabular*}{\linewidth}{@{\extracolsep{\fill}}lcccc@{\extracolsep{\fill}}}
\toprule
$M$ & $\epsilon_{frame}$& $\epsilon_{acc.}$   & $\epsilon_{dice}$ &  $\epsilon_{drift}$ \\
\midrule

   20   & $0.47\pm0.04$ &    $20.96\pm1.35$ &  $0.39\pm0.09$ &   $30.03\pm2.07$   \\ 
   49   & $0.45\pm0.08$ &    $21.77\pm2.90$ &  $0.40\pm0.11$ &   $32.41\pm3.83$    \\
   75   & $0.39\pm0.08$ &    $20.07\pm3.81$ &  $0.39\pm0.13$ &   $29.34\pm5.32$    \\
  100   & $0.32\pm0.07$ &    $13.97\pm4.88$ &  $0.24\pm0.13$ &   $18.17\pm6.01$  \\

\bottomrule
\end{tabular*}
\end{center}
\end{table}
\noindent

\balance

\begin{sidewaysfigure*}[hbt!]%
\renewcommand{\thefigure}{S-\arabic{figure}}
\centering
\includegraphics[width=\textwidth]{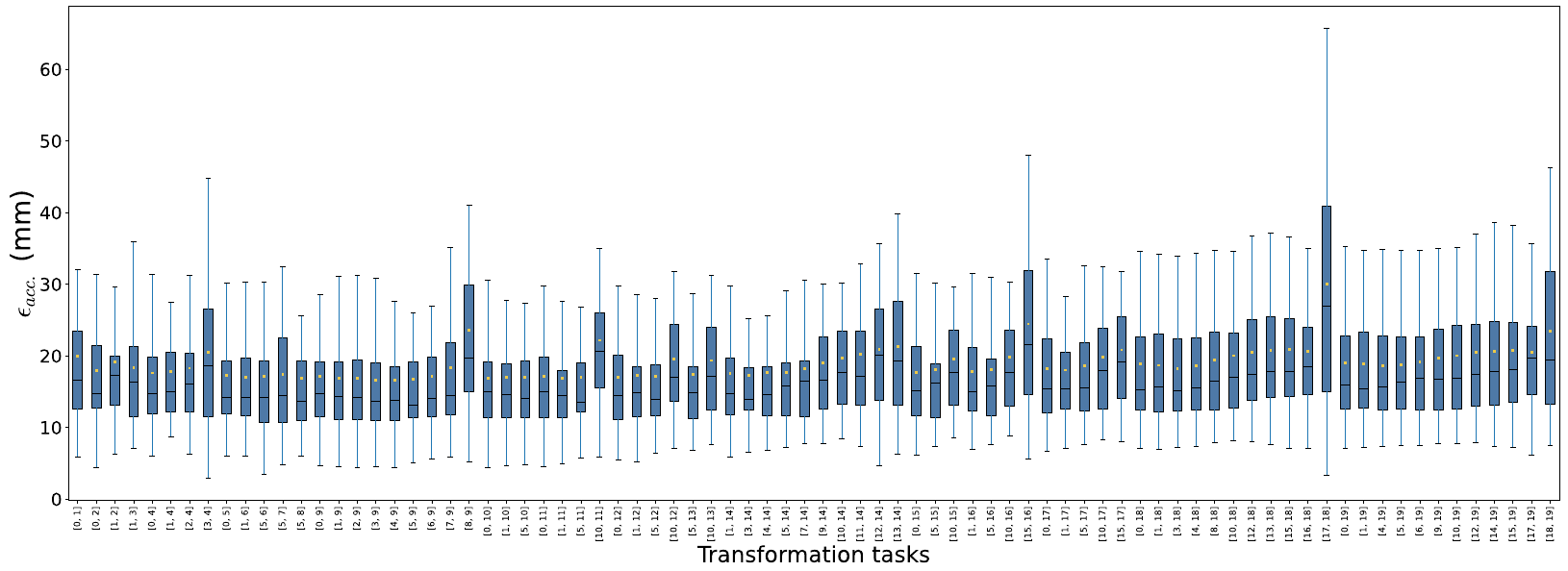}
\caption{Performance of various transformation tasks in terms of $\epsilon_{acc.}$ when sequence length $M$ = 20. Each bar denotes the distribution of $\epsilon_{acc.}$ using the specific transformation $T_{j\leftarrow i}$ over all scans in the test set.} \label{acc_err_m20-s}
\end{sidewaysfigure*}

\begin{sidewaysfigure*}[hbt!]%
\renewcommand{\thefigure}{S-\arabic{figure}}
\centering
\includegraphics[width=\textwidth]{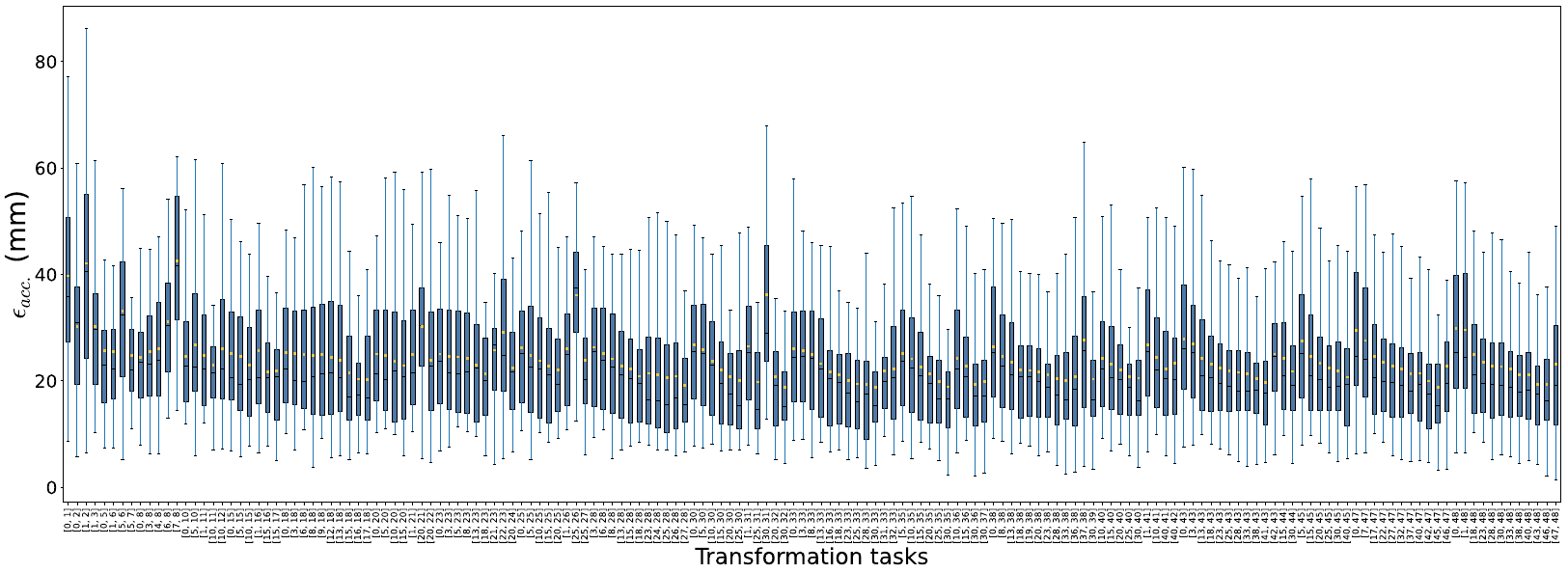}
\caption{Performance of various transformation tasks in terms of $\epsilon_{acc.}$ when sequence length $M$ = 49. Each bar denotes the distribution of $\epsilon_{acc.}$ using the specific transformation $T_{j\leftarrow i}$ over all scans in the test set.}\label{acc_err_m49-s}
\end{sidewaysfigure*}

\begin{sidewaysfigure*}[hbt!]%
\renewcommand{\thefigure}{S-\arabic{figure}}
\centering
\includegraphics[width=\textwidth]{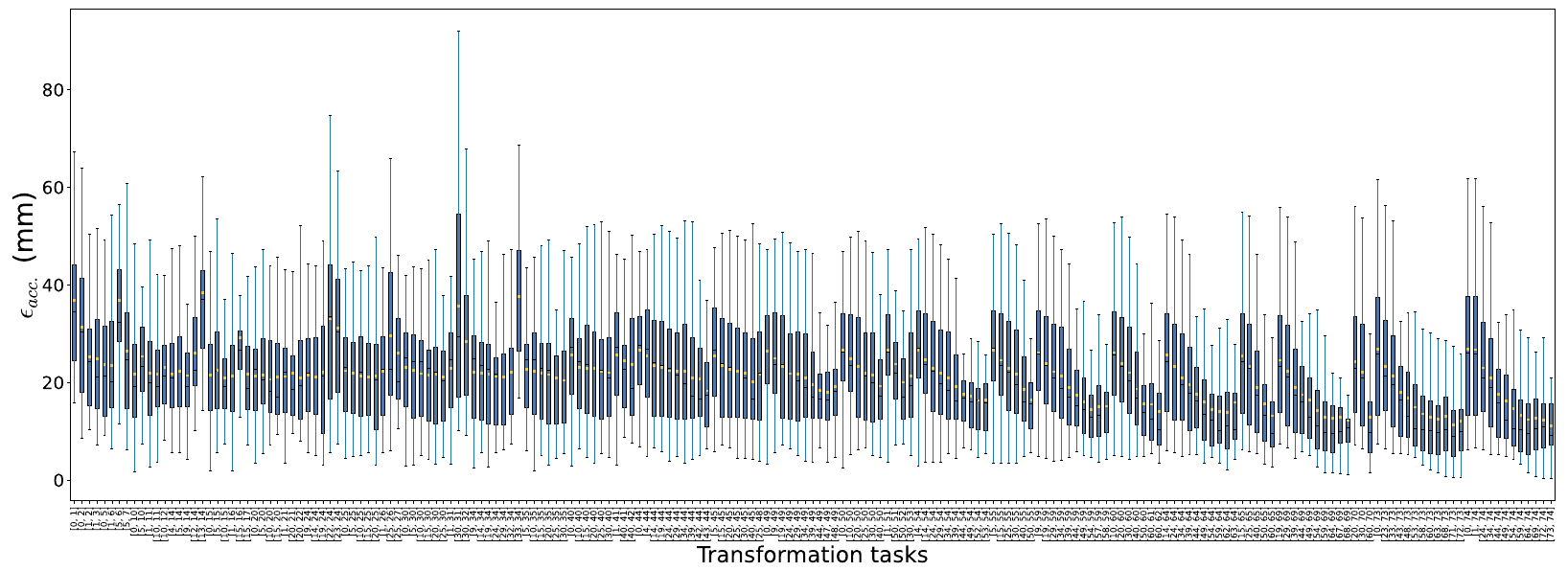}
\caption{Performance of various transformation tasks in terms of $\epsilon_{acc.}$ when sequence length $M$ = 75. Each bar denotes the distribution of $\epsilon_{acc.}$ using the specific transformation $T_{j\leftarrow i}$ over all scans in the test set.}\label{acc_err_m75-s}
\end{sidewaysfigure*}

\begin{sidewaysfigure*}[hbt!]%
\renewcommand{\thefigure}{S-\arabic{figure}}
\centering
\includegraphics[width=\textwidth]{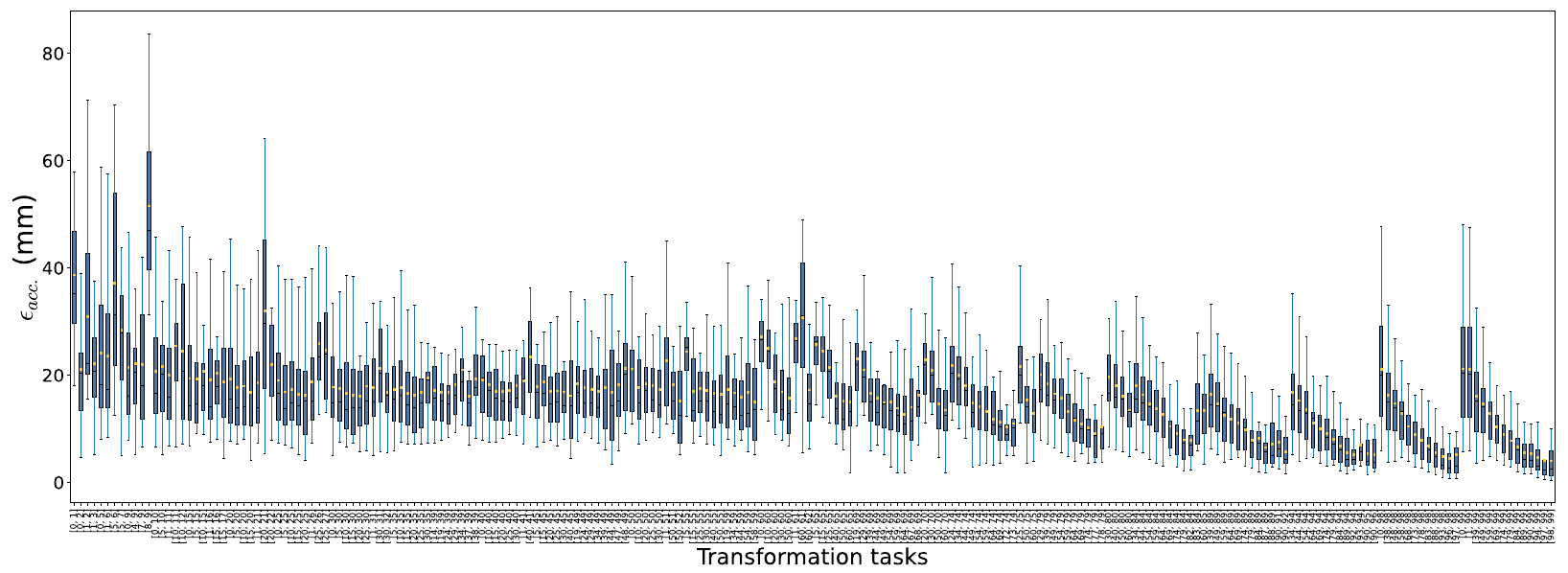}
\caption{Performance of various transformation tasks in terms of $\epsilon_{acc.}$ when sequence length $M$ = 100. Each bar denotes the distribution of $\epsilon_{acc.}$ using the specific transformation $T_{j\leftarrow i}$ over all scans in the test set.}\label{acc_err_m100-s}
\end{sidewaysfigure*}

\end{document}